\documentclass[apj,numberedappendix]{emulateapj}
\usepackage{epsfig}

\newcommand{\rxjw}{RX~J1856.5$-$3754}
\newcommand{\rxj}{RX~J0720.4$-$3125}
\defcitealias{kvka02}{KvKA02}
\defcitealias{ak04}{AK04}
\let\mc\multicolumn
\let\gsim\gtrsim
\let\lsim\lesssim
\newcommand{\parsym}{\ensuremath{\pi}}
\newcommand{\mstis}{\ensuremath{m_{50}}}
\newcommand{\mhrcb}{\ensuremath{m_{475}}}
\newcommand{\mb}{\ensuremath{B}}
\newcommand{\mr}{\ensuremath{R}}
\newcommand{\bmr}{\ensuremath{B-R}}
\newcommand{\hrcbmr}{\ensuremath{m_{475}-R}}
\newcommand{\hrcbmstis}{\ensuremath{m_{475}-m_{50}}}
\newcommand{\stismr}{\ensuremath{m_{50}-R}}

\newcommand{\expnt}[2]{\ensuremath{#1 \times 10^{#2}}}   
\newcommand{\hst}{\textit{HST}}
\newcommand{\rosat}{\textit{ROSAT}}

\newcommand{\syst}{0.4}
\newcommand{\pmsyst}{1}

\begin{document}

\shorttitle{The Distance of RX J0720.4$-$3125}
\shortauthors{Kaplan et~al.}

\title{The Distance to the Isolated Neutron Star \rxj}

\author{D.~L.~Kaplan\altaffilmark{1}, M.~H.~van
  Kerkwijk\altaffilmark{2}, and J.~Anderson\altaffilmark{3}}

\altaffiltext{1}{Pappalardo Fellow; Kavli Institute for Astrophysics and Space
  Research and Department of Physics, Massachusetts Institute of
  Technology, Cambridge, MA 02139; dlk@space.mit.edu.} 
\altaffiltext{2}{Department of Astronomy and Astrophysics, University
  of Toronto, 50 St.\ George Street, Toronto, ON M5S 3H4, Canada;
mhvk@astro.utoronto.ca.}
\altaffiltext{3}{Department of Physics and Astronomy, MS 61, Rice
  University, P.O.\ Box 1892, Houston, TX 77251; jay@eeyore.rice.edu} 

\slugcomment{To be published in ApJ}

\begin{abstract}
We have used a set of dedicated astrometric data from the
\textit{Hubble Space Telescope} to measure the parallax and proper
motion of the nearby neutron star \rxj.  At each of eight epochs
over two years, we used the High Resolution Camera of the Advanced
Camera for Surveys to measure the position of the $B=26.6$ target to a
precision of $\sim\!2$~mas ($\sim\!0.07$~pix) relative to 22 other
stars.  From these data we measure a parallax of $\parsym=2.8\pm0.9$~mas
(for a distance of $360^{+170}_{-90}$~pc) and a proper motion of
$\mu=107.8\pm1.2\mbox{ mas yr}^{-1}$.  Exhaustive testing of every
stage of our analysis suggests that it is robust, with a maximum
systematic uncertainty on the parallax of \syst~mas.  The distance is
compatible with earlier estimates made from scaling the optical
emission of \rxj\ relative to the even closer neutron star \rxjw.  The
distance and proper motion imply a transverse velocity of
$180^{+90}_{-40}\mbox{ km s}^{-1}$, comparable to velocities observed
for radio pulsars.  The speed and direction suggest an origin for
\rxj\ in the Trumpler~10 OB association $\sim\!0.7$~Myr ago, with a
possible range of 0.5--1.0~Myr given by the uncertainty in the
distance.
\end{abstract}


\section{Introduction}
One of the many interesting results from \rosat\ All-Sky Survey
\citep{rbs} was the discovery of seven objects that appear to be
nearby, thermally-emitting neutron stars that have little if any
magnetospheric emission (for recent reviews, see
\citealt{haberl04,haberl06,vkk06}).  These objects, known most
commonly as ``isolated neutron stars,'' (INS) are distinguished by
their long spin periods ($\gsim\!3$~s, when measured), largely thermal
spectra with cool temperatures ($kT \lsim 100$~eV), faint optical
counterparts, and lack of radio emission.

Thermally-emitting neutron stars have been the targets of many
observations, as they can potentially be used to constrain the
equation of state (EOS) of neutron stars, and thereby explore nuclear
physics in realms inaccessible from laboratories \citep[e.g.,][]{lp00}.  Two
main approaches are used.  The first, using the spectrum, seems
simple: determine the effective angular size from spectral fits,
multiply by the distance (obtained by other means), and one has the
apparent radius.  This radius can be converted into the physical
radius through use of mass.  The radius is the crucial quantity in
differentiating between EOS, as most EOS predict a distinctive but
small range of radii for a large range of masses.  The second method
is to use measurements of an ensemble of neutron stars to constrain
cooling curves, which are themselves sensitive to the EOS and the
interior composition \citep[for a review,][]{yp04}.  This also seems
simple: measure the received flux of an object, convert to a
luminosity using the distance, and with the age place that object on a
cooling diagram.  The closest of the INS, \object[RX
J1856.5-3754]{\rxjw}, has been the subject of much inquiry  for
just these purposes \citep[e.g.,][]{wl02,dmd+02,br02}.

\begin{deluxetable}{l l c c c}
\tablecaption{\hst\ ACS/HRC F475W Observation Summary\label{tab:obs}}
\tablewidth{0pt}
\tablehead{
&\mc{2}{c}{\dotfill Date\dotfill} &
\colhead{Orientation\tablenotemark{a}} & \colhead{\mhrcb\tablenotemark{b}}\\ 
\colhead{Epoch} & \colhead{(UT)} & \colhead{(MJD)} & \colhead{(deg)}& \colhead{(mag)}
}
\startdata
1 & 2002~Jul~04 & 52460.0 &353.0108 &26.66\\
2 & 2002~Sep~15 & 52532.6 &\phn72.5457&26.54 \\
3 & 2003~Jan~06 & 52645.1 &175.9862   &26.45 \\
4 & 2003~Mar~16 & 52714.0 &254.8005   &26.57 \\
5 & 2003~Aug~04 & 52855.1 &\phn28.0808&26.39 \\
6 & 2003~Sep~21 & 52903.0 &\phn77.1792&26.45 \\
7 & 2004~Jan~07 & 53011.9 &178.4808   &26.42 \\
8 & 2004~Mar~19 & 53083.1 &258.1951   &26.64 \\
\enddata
\tablecomments{For each epoch, eight exposures were taken during two
  orbits.  For epochs 1--4, the integration times for exposures 1--4
  were 600~s, while those for exposures 5--8 were 630~s.  For epochs
  5--8, the exposure times were 625~s in the first orbit, and 660~s in
  the second.  The dither offsets along the $x$ and $y$ axes of the
  detector (written in base eight, where, e.g., 2.2 is
  $2+\frac{2}{8}$), were [0.0, 0.0], [2.2, 2.2], [4.4, 4.4], [6.6,
  6.6], [1.1, 5.5], [3.3, 7.7], [5.5, 1.1], and [7.7, 3.3]).}
\tablenotetext{a}{This is the  value of the \texttt{orientat}
  keyword, roughly corresponding to
  the $y$-axis on the HRC.  The transformation angles in
  Table~\ref{tab:epoch} are relative to these.}
\tablenotetext{b}{The F475W magnitude of \rxj, measured at each
  epoch.  The uncertainty on each measurement is $\sim\!0.09$~mag.}
\end{deluxetable}

\begin{figure*}
\plotone{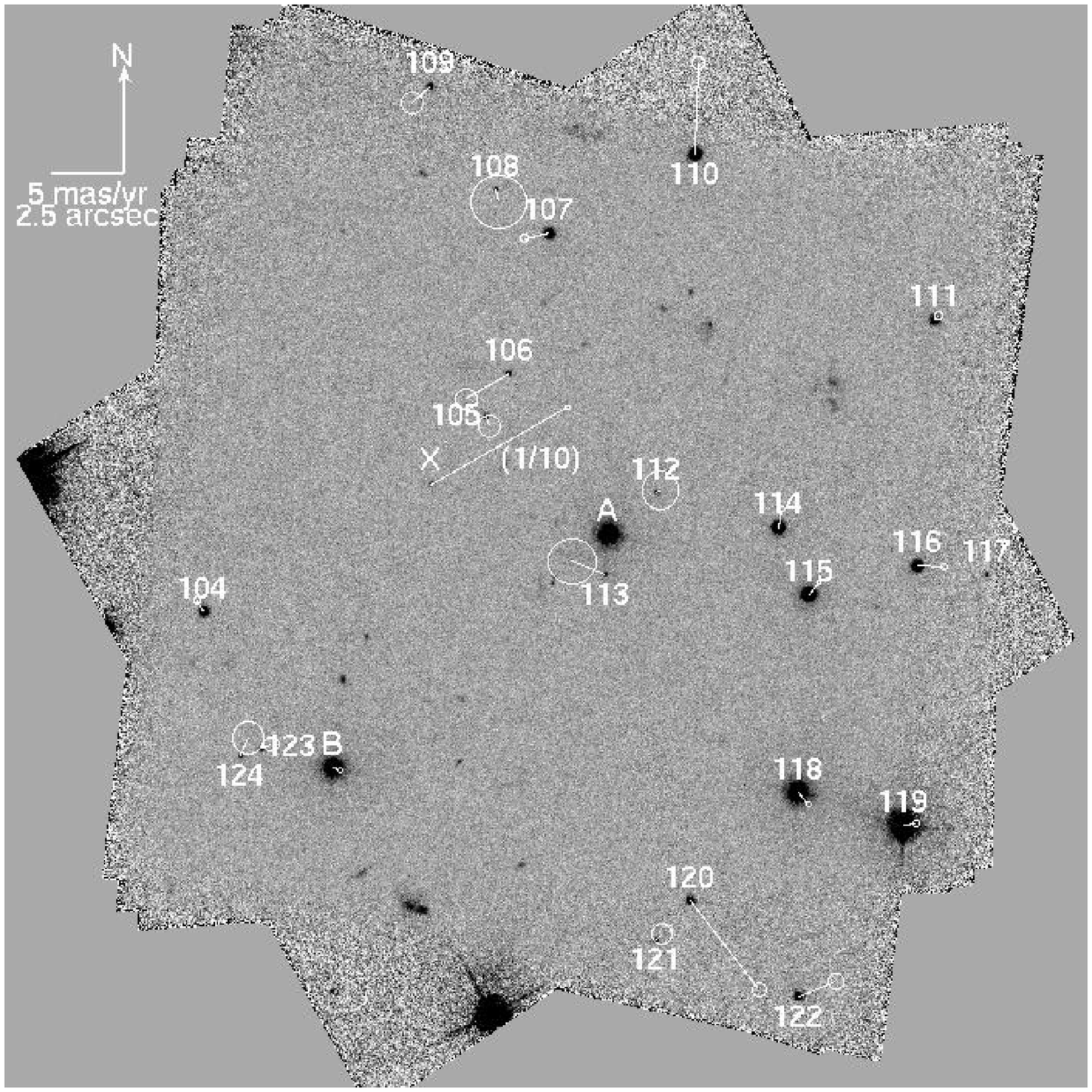}
\caption{Combined image of all 64 HRC exposures of the field of \rxj\
  (made using \texttt{multidrizzle}; \citealt{kfhh02}).  The objects
  used for the astrometry are labeled (remaining objects were either
  too faint or too extended to be used).  We also show the fitted
  proper motions of those objects, with the ellipses indicating the
  1$\sigma$ uncertainties on the proper motion.  Note that the proper
  motion of \rxj\ (object ``X'') has been scaled down by a factor of $1/10$ compared
  to the other objects in the field.  The scales and orientation are
  indicated in the upper left corner.}
\label{fig:hrc}
\end{figure*}

While the basic approaches --- inferring radii from broad-band
spectral fits and comparing cooling luminosities to models --- are
easily stated, complications arise from uncertainties in distances and
ages; additional complications may result from the presence of strong
surface magnetic fields and unknown surface compositions.  Progress on
these last issues has been made recently from X-ray spectroscopy
\citep{hsh+03,vkkd+04,hmz+04,zct+05} and timing \citep{kvk05,kvk05b},
but distances and ages are difficult to measure.  The technique
necessary for both is high-precision optical astrometry, as shown by
the example of \rxjw: with a series of \textit{Hubble Space Telescope}
(\hst) observations it was possible to measure the parallax
(\citealt*[][ hereafter \citetalias{kvka02}]{kvka02}; \citealt{wl02}),
and the proper motion traced the object to an OB association where it
was likely born \citep{wal01}.

While \rxjw\ is the brightest and closest of the INS, and has thus
garnered most of the attention, it is best to try to use multiple
sources.  This especially since each source appears to have
peculiarities, be it stronger or weaker timing noise, stronger or
weaker features in the X-ray spectra, presence or absence of long-term
variations, or the presence of an H$\alpha$ nebulae (in the case of
\rxjw; \citealt{vkk01b}).  Likely, secure results will only be
obtained if we understand and can correct for these differences.
Here, we discuss high-precision astrometric observations of the second
brightest source, \object[RX J0720.4-3125]{\rxj}.

\rxj\ was discovered by \citet{hmb+97} as a soft ($kT\simeq80$~eV),
bright X-ray source in the \rosat\ All-Sky Survey.  Given its low
hydrogen column density ($N_H\sim \expnt{1}{20}\mbox{~cm}^{-2}$),
nearly sinusoidal 8.39-s pulsations, relatively constant X-ray flux,
and faint ($B=26.6$~mag), blue optical counterpart \citep{kvk98,mh98},
it was classified as a nearby, isolated, thermally-emitting neutron
star.  X-ray timing observations \citep{kvk05} give a characteristic
age of 2~Myr and a magnetic field strength of $\expnt{2.4}{13}$~G,
consistent with suggestions of the source being an off-beam,
moderately strong-field radio pulsar \citep{zhc+02,kkvkm02}.  While
the above properties make \rxj\ a prototypical INS, X-ray monitoring
over the last few years has uncovered unique behavior: the X-ray
spectrum and pulse shape have been evolving \citep{dvvmv04,vdvmv04}.
This may indicate free precession \citep{htdv+06}, although this
interpretation is not unique \citep{vkk06}.  Scaling to \rxjw, its
distance was estimated to be $\sim\!300$~pc \citepalias{kvka02}.
\citet{msh+05} measured a proper motion of $97\pm12\mbox{ mas
yr}^{-1}$ from ground-based astrometry, and suggested the source might
originate in the OB association Trumpler 10, something we
independently derived from part of the data presented here
\citep{kaplan04}.

In this paper, we present a detailed analysis of \hst\ observations of
\rxj\ specifically designed for astrometry, which we use to measure
its proper motion and parallax.  The organization of this paper is as
follows.  First, in \S~\ref{sec:obs}, we present the astrometric
observations and discuss additional observations we used for
photometry and for tying our astrometry to the International
Coordinate Reference System (ICRS).  We also discuss the basic
reduction of these data.  Then, in \S~\ref{sec:photpar}, we use the
photometry to determine photometric parallaxes of the reference stars,
which we use in \S~\ref{sec:parallax} for the parallax.  We discuss
the implications in \S~\ref{sec:disc} and conclude in
\S~\ref{sec:conc}.

Our analysis is complex and necessarily involves choices among
alternate schemes.  Previous attempts to measure parallaxes with \hst\
made choices that were later called into question (\citealt{wal01};
\citetalias{kvka02}; \citealt{wl02}).  To clarify our analysis, and to
set out the framework for future work, we give full details on the
parallax measurement in App.~\ref{app:ast}.  We then show that our
results are robust to the choices we made by describing various
alternate analyses in App.~\ref{app:alt}.  In what follows, we define
our proper motions in Right Ascension and Declination
$(\mu_\alpha,\mu_\delta)$ such that the scales are the same and no
$\cos\delta$ term is necessary.  All uncertainties are $1\sigma$
unless otherwise indicated.  In addition to the results presented in
Tables~\ref{tab:ast}--\ref{tab:par}, we make the raw data available
electronically.

\section{Observations, Reduction, and Analysis}
\label{sec:obs}

Our main data set consist of 64 exposures taken during eight visits of
the field of \rxj\ with \hst; a summary is given in
Table~\ref{tab:obs}.  All data were taken with the High Resolution
Camera (HRC, with a plate-scale of $28.27{\rm~mas\,pix}^{-1}$) and the
F475W (SDSS $g^\prime$) filter, which has good sensitivity to the blue
colors of \rxj.

We tried to optimize our observations for measuring an accurate
parallax in three ways.  First, we observed over two years, to
minimize the covariance between proper motion and parallax, and verify
repeatability.  Second, we observed four times per year rather than
just at the parallactic maxima, to reduce possible systematic effects
of observing at orientations different by 180\arcdeg\ only.  Third,
for the eight exposures in each epoch, we chose dither positions
optimized for sampling fractional pixel phase.

\begin{figure}
\plotone{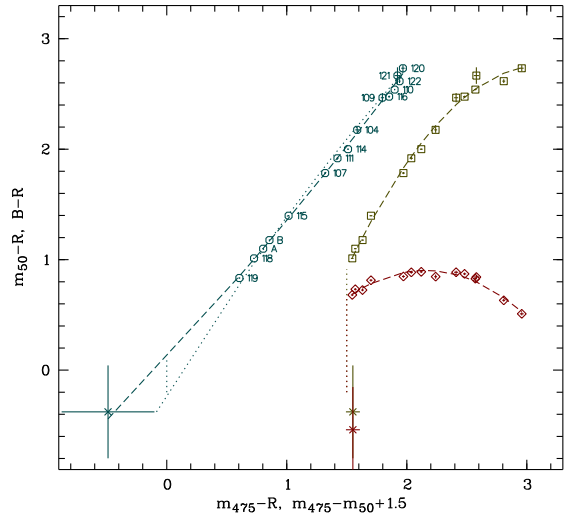}
\caption{Color-color relations used to infer \bmr\ and \mr\ for stars
  for which the ground-based photometry was inaccurate.  Shown are
  relations between \bmr\ and \hrcbmr\ (circles), \bmr\ and \hrcbmstis\
  (squares), and \stismr\ and \hrcbmstis\ (lozenges),
  with the \hrcbmstis\ magnitudes offset by 1.5~mag for clarity;
  the zero points are indicated vertical dotted lines.  Overdrawn
  (long-dashed lines) are the empirical quadratic fits from
  Table~\ref{tab:colors} to all points
  except \rxj\ (crossed points), as well as the relation between \bmr\
  and \hrcbmr\ expected from synthetic photometry (dotted line;
  \citealt{sjb+05}).} 
\label{fig:colors}
\end{figure}

In our analysis, we also use three sets of observations presented
previously.  For photometry, these are B- and R-band images taken with
the Low Resolution Imaging Spectrometer (LRIS) on Keck \citep{kvk98},
and images taken with the Space Telescope Imaging Spectrograph (STIS)
on \hst\ \citep{kvkm+03}, and for our astrometric tie, B-band
observations taking with the Focal Reducer, low-dispersion
Spectrograph (FORS1) on the Very Large Telescope \citep*{mzh03}.

Below, we describe first the reduction of all data sets, and then the
way we measured positions from the HRC images, tied our astrometry to
the International Celestial Reference System, and obtained photometry
from the HRC, STIS and LRIS data.  For reference, we show a combined
image of all HRC data in Fig.~\ref{fig:hrc}, with stars used in our
analysis labeled.

\subsection{Reduction}
\label{sec:reduction}

For each data set, the basic reduction was standard: bias and dark
current were subtracted, and possible pixel-to-pixel sensitivity
variations were corrected for using flat fields.  For the \hst\ data,
these steps are taken in the standard pipelines, while for the LRIS
data they were performed by \citet{kvk98}.  For the FORS images, which
we retrieved from the archive and re-analyzed, we determined the bias
level from the overscan regions and constructed a flat field from dawn
sky images.

\begin{deluxetable*}{lllllllccl}
\def\bmr{{\it B},{\it R}}
\def\hrcbmr{{\it 475},{\it R}}
\def\hrcbmstis{{\it 475},{\it 50}}
\tablecaption{Photometry and photometric parallaxes for stars in the
  field of \rxj\label{tab:photpar}}
\tabletypesize{\footnotesize}
\tablewidth{0pt}
\tablehead{&\multicolumn{4}{c}{\dotfill Observed\dotfill}&
\multicolumn{5}{c}{\dotfill Inferred\dotfill}\\
&\colhead{\mb}&\colhead{\mhrcb}&
\colhead{\mstis}&\colhead{\mr}&
\colhead{$\mr_{\rm ad}$}&\colhead{$(\bmr)_{\rm ad}$}&&
\colhead{$M_{R,\rm MS}$}&\colhead{$\parsym_{\rm phot}$}\\
\colhead{ID}&
\colhead{(mag)}&\colhead{(mag)}&\colhead{(mag)}&\colhead{(mag)}&
\colhead{(mag)}&\colhead{(mag)}&\colhead{Source}&\colhead{(mag)}&
\colhead{(mas)}
}

\startdata
X&26.62(17)&26.51(4)&26.45(5)&27.0(4)&27.0(4)&$-$0.4(4)&\bmr&\nodata&\nodata\\
A&20.517(4)&20.221(5)&20.150(10)&19.418(3)&19.418(3)&1.099(5)&\bmr&4.5&0.105\\
B&20.789(4)&20.468(5)&20.337(10)&19.612(3)&19.612(3)&1.177(5)&\bmr&4.8&0.110\\
104&23.60(2)&23.012(5)&22.272(11)&21.425(15)&21.425(15)&2.17(3)&\bmr&7.5&0.166\\
105&27.0(2)&25.801(18)&24.578(15)&23.84(2)&23.84(2)&2.68(5)&\hrcbmr&9.5&0.134\\
106&26.47(13)&25.653(18)&24.391(14)&23.66(2)&23.66(2)&2.73(5)&\hrcbmr&9.7&0.165\\
107&23.384(9)&22.920(6)&22.448(11)&21.600(4)&21.600(4)&1.784(10)&\bmr&6.6&0.101\\
108&\nodata&26.77(4)&25.50(2)&24.98(8)&24.82(9)&2.67(8)&\hrcbmstis&9.4&0.082\\
109&25.08(4)&24.410(12)&23.499(12)&22.613(7)&22.613(7)&2.44(3)&\hrcbmr&8.3&0.140\\
110&22.735(5)&22.094(3)&21.022(10)&20.195(3)&20.195(3)&2.540(6)&\bmr&8.7&0.51\\
111&23.731(12)&23.235(8)&22.698(11)&21.813(4)&21.813(4)&1.918(12)&\bmr&6.9&0.105\\
112&\nodata&26.43(3)&25.127(20)&24.4(2)&24.47(7)&2.69(7)&\hrcbmstis&9.5&0.102\\
113&\nodata&26.79(4)&25.60(3)&\nodata&24.86(9)&2.62(9)&\hrcbmstis&9.1&0.072\\
114&22.514(6)&22.024(5)&21.404(10)&20.514(3)&20.514(3)&2.000(6)&\bmr&7.1&0.21\\
115&21.774(4)&21.392(6)&21.190(10)&20.377(3)&20.377(3)&1.398(5)&\bmr&5.6&0.111\\
116&23.045(8)&22.422(8)&21.441(10)&20.569(4)&20.569(4)&2.476(9)&\bmr&8.5&0.38\\
117&27.1(3)&26.43(3)&25.30(2)&24.48(6)&24.53(7)&2.59(7)&\hrcbmstis&9.0&0.077\\
118&20.510(4)&20.225(5)&20.179(10)&19.498(3)&19.498(3)&1.012(5)&\bmr&4.2&0.087\\
119&19.689(3)&19.461(7)&\nodata&18.855(3)&18.855(3)&0.834(4)&\bmr&3.5&0.087\\
120&24.73(3)&23.965(11)&22.507(11)&21.997(8)&21.997(8)&2.73(3)&\bmr&9.8&0.36\\
121&25.86(7)&25.115(17)&24.036(13)&23.193(15)&23.193(15)&2.62(4)&\hrcbmr&9.1&0.154\\
122&24.135(16)&23.459(5)&22.151(11)&21.519(6)&21.519(6)&2.616(17)&\bmr&9.1&0.33\\
123&25.92(16)&25.229(14)&23.861(12)&23.22(3)&23.25(5)&2.71(5)&\hrcbmstis&9.7&0.191\\
124&27.3(3)&26.304(20)&24.759(16)&24.56(7)&24.31(6)&2.75(6)&\hrcbmstis&9.9&0.132\\

\enddata

\tablecomments{{\em Observed:} All errors quoted exclude the
  zero-point uncertainties to which the measurements are on the Vega
  system.  These are $\lesssim\!0.02~$mag for all but \mstis, which is
  only roughly calibrated (see \S\ref{sec:photometry}).  Reasons for
  missing measurements are: 108 and 112 (\mb): too faint; 113 (\mb\
  and \mr): too close to star A; 119 (\mstis): overexposed.  {\em
  Inferred:} $R_{\rm ad}$ and $(\bmr)_{\rm ad}$ are our adapted values,
  inferred from the magnitudes listed under Source.  The inferred
  absolute magnitudes $M_{R,\rm MS}$ have uncertainties around
  0.4~mag, and the corresponding uncertainties in the photometric
  parallaxes $\parsym_{\rm phot}$ are about 20\% (see
  \S\ref{sec:photpar}).}
\end{deluxetable*}

\begin{deluxetable*}{lclllc}
\tablecaption{Color transformations used to infer \bmr\ and $R$ for stars in
  the field.\label{tab:colors}}
\tablewidth{0pt}
\tablehead{%
&&&&&\colhead{rms}\\
\colhead{Color}&&\mc{3}{c}{Empirical Relation}&\colhead{(mag)}
}
\startdata
$\bmr$&$=$&$+0.116$&$+1.173\,(\hrcbmr)$&$+0.067\,(\hrcbmr)^2$&$+0.030$\\
$\bmr$&$=$&$+0.915$&$+2.270\,(\hrcbmstis)$&$-0.700\,(\hrcbmstis)^2$&$+0.049$\\
$\stismr$&$=$&$+0.654$&$+0.759\,(\hrcbmstis)$&$-0.579\,(\hrcbmstis)^2$&$+0.028$\\

\enddata
\end{deluxetable*}

In preparation for further analysis, we flagged bad pixels and made
averages.  For the HRC data, we used a procedure similar to that of
{\tt multidrizzle} \citep{kfhh02}, but which does not correct for
distortion.  First, we removed pixels flagged as bad in the data
quality array, as well as two obvious bad columns [upwards from pixel
$(x,y)=(200,548)$ and $(617,228)$].  Next, we subtracted the sky level
and resampled on an eight times finer grid, for which the dither
positions correspond to integer pixel offsets.  Using these offsets,
we registered the images for each epoch and constructed an initial
guess at a cosmic-ray free, average image from the two lowest values
at each position.

For each image, we flagged as cosmic-ray hits all pixels in excess by
more than $5\sigma$ of those in the initial average (resampled back to
the original resolution) as well as all adjacent pixels.  Here, we
estimated $\sigma$ by adding, in quadrature, the read-out noise, the
Poisson noise due to sky and signal (as inferred from the average), as
well as terms equal to 5\% of the signal and 5\% of the range in
signal in the surrounding $3\times3$ pixel box.  The latter terms
ensure centers of bright stars are not incorrectly flagged as bad due
to registration errors, focus changes, etc.  We also constructed a
final average for each epoch from those cleaned images.

For the STIS data, the analysis was similar, except that no resampling
was necessary, since integer pixel offsets were used.  For the LRIS
and FORS images, we used more standard methods to correct for
cosmic-ray hits, in which they are identified by their narrow width
compared to the seeing, and replaced by interpolation.  For the
photometry and astrometry below, averages were made of these filtered
images, registered to integer pixel shifts.

\subsection{Position Measurements}
\label{sec:hrc}
We determined positions using the ``effective'' point-spread function
(ePSF) technique of \citet{ak00}.  We use the ePSF determined for the
HRC with the F475W filter by \citet[ hereafter
\citetalias{ak04}]{ak04}, and use a variation of their procedure for
the fits.  In App.~\ref{app:pos}, we describe where we deviate and
discuss further variations; fortunately, our results do not depend
much on our choices, although the method described here gives slightly
superior precision.

\begin{figure}
\plotone{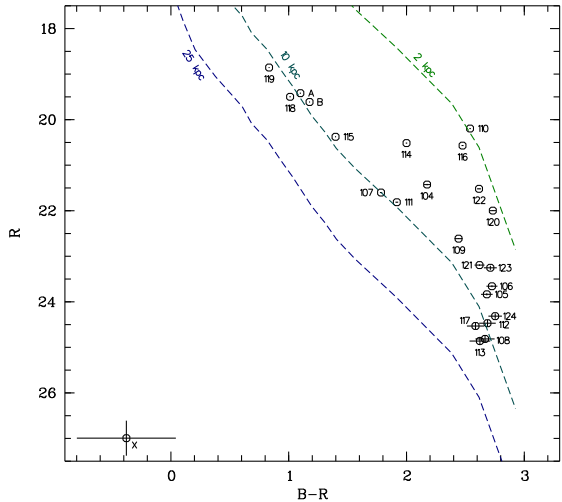}
\caption{Color-magnitude diagram, showing $R$ magnitude versus \bmr\
  color for the objects in Table~\ref{tab:photpar}.  We also show main
  sequences with reddening $A_V=0.25$~mag at 2, 8.5, and 25~kpc (as
  labeled).  One sees that most stars are at about 10~kpc distance,
  implying that their photometric parallaxes are small, about
  0.1~mas.}
\label{fig:hrd}
\end{figure}

In our procedure, we fit stellar images on each exposure using the
pixels in a radius of 2.56 pixels around an initial guess for the
centroid (determined from a Gaussian fit in the average image),
weighting data points with uncertainties
$\sigma=(\textrm{RN}^2+\max(F,\textrm{sky})/\textrm{G})^{1/2}$ (where
$\textrm{RN}=6$~DN is the read-noise, which we determined empirically
from variations in background regions, and
$\textrm{G}=1{\rm~e^-\,DN^{-1}}$ is the gain for pipe-line processed
HRC images).  We fit simultaneously for the position and amplitude of
each star, but fix the sky to the header value {\tt mdrizsky}, which
is determined by the pipeline drizzle process (and is very similar to
the median or mode of all pixel values).  Finally, we correct the
positions for distortion in the HRC using the solution from
\citetalias{ak04}.

To estimate uncertainties on the positions, we used the standard
technique for $\chi^2$ fitting in which we determined the $x$ and $y$
offsets at which $\chi^2$ increased by~1.  It is worth noting that,
generally, this is only valid if the fit is acceptable, i.e., if
$\chi^2$ roughly equals the number of degrees of freedom.  For the
brightest stars, we often find much larger $\chi^2$, since the PSF
varies due to focus changes, etc.  One could rescale the errors, but
this would lead one to overestimate the uncertainties of positions
of stars relative to each other (which is what enters our analysis),
since PSF changes affect all stars in the same way.  Therefore, we
retained the formal uncertainties, verifying that they lead to
acceptable fits (see below and App.~\ref{app:uncert}).

With positions and uncertainties in hand, we solve for the best-fit
transformation between the first and the other seven exposures for
each epoch (for details, see App.~\ref{app:ast}), and determine one
set of average, distortion-corrected positions per epoch.  Given our
weighting, the four brightest stars (A, B, 118, and 119) dominate the
fits.  For the transformation, we follow \citetalias{ak04} and use a
6-parameter, bi-linear transformation, as we do for the transformation
between epochs (see \S~\ref{sec:parallax}).  This fits for a central
position, the overall plate-scale and rotation of the image, and a
second plate-scale and rotation that reflect differences in scale and
orientation between the two axes (we test variations on this in
App.~\ref{app:combination}).

In general, when we included all measurements, the fit quality for
some transformations was very poor.  Inspecting the outliers, we found
that within the fitted region of most there were some pixels flagged
as bad.  These pixels are not included in the ePSF fit, so to first
order they should not affect the result.  There is a second-order
effect, however: that arises because our model ePSF is not a perfect
match to the real one.  As stated above, generally, this does not
matter for relative positions, since the ePSF is wrong in the same way
for all stars.  But this relies on the same part of the ePSF being
used, which is not possible if bad pixels are present.  Because of
this, we decided to exclude all position measurements that had bad
pixels.  In addition, we excluded all measurements of star 114 in
epoch~4, where it is close to the HRC occulting finger and clearly
deviant, and three further significant outliers without obvious causes
(star 115 in epoch~2, exposure~2; 118 in 3,~5; and A in 7,~7).

With the above exclusions, the fits are fairly good, with the overall
$\chi^2_{\rm red}$ ranging from 1.00 for epoch~8 to 1.48 for epoch~5.
Furthermore, all individual sources were found to fit well too (for
details, including simulated data, see App.~\ref{app:uncert} and
Fig.~\ref{fig:err}).

\subsection{Absolute Astrometry}
\label{sec:absast}
While the HRC observations provide very accurate relative positions,
the precision with which these can be tied to the International
Celestial Reference System (ICRS) is not as good, as it is determined
by the less accurately known positions of the guide stars used. To
improve the precision, we use ground-based observations to tie our
measurements to the UCAC2 catalog \citep{zuz+04}, which is currently
the most precise representation of the ICRS for stars fainter than
those measured by Hipparcos.

We first tied a short FORS image taken on 31~December~2002, which is
very close in time to our third HRC observation, to the UCAC2 catalog
(corrected to the same epoch using the UCAC2 proper motions).  We
measured centroids of 13 UCAC2 stars present on the image and
corrected for radial distortion using the quadratic relation given by
\citet*{jobs06}.  A four-parameter transformation sufficed, with the
offset accurate to 10~mas, and scale and rotation different from
nominal by $-0.028\pm0.012$\% and $-0\fdg076\pm0\fdg008$,
respectively.  The fit was good, with the residuals consistent with
the UCAC2 uncertainties.

We then used 41 stars to transfer the tie to a deeper image,
consisting of twenty 620-s images from 29, 30, and 31 December~2002.
This fit was again good, and the uncertainties negligible compared to
the first step.  Finally, we tied our astrometry to the HRC using 18
stars for which we could obtain accurate FORS positions.  Using HRC
positions evaluated at the FORS epoch using the proper motions derived
in \S\ref{sec:parallax}, we find an excellent tie, with offsets
accurate to $\sim\!2$~mas and a scale and position angle different
from the nominal values by $-0.034\pm0.015\%$ and
$-0\fdg018\pm0\fdg010$, respectively (here, the uncertainties take
into account that the HRC shows deviations from equality of scales and
orthogonality; see \S~\ref{sec:parallax}).

The overall precision with which our measurements are on the UCAC2
system is limited by the first step, and is $\sim\!11$~mas, similar to
the precision with which UCAC2 is on the ICRS \citep{zuz+04}.  For the
scale and position angle, the first and last step contribute; we
estimate that the scale is accurate to 0.02\% and the position angle
to $0\fdg011$.

Below, we will give positions relative to star A, fixing its position
to $\alpha_{\rm J2000}=07^{\rm h}20^{\rm m}24\fs4837$, $\delta_{\rm
J2000}=-31^\circ25'51\farcs786$.  We will also correct the nominal
scale and position angles of epochs 3 and~7 using the values found
above.

\begin{figure}[t]
\plotone{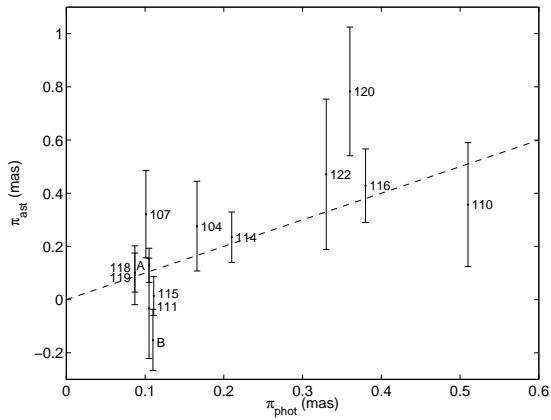}
\caption{Photometric versus astrometric parallaxes for the stars from
  Table~\ref{tab:photpar} for which the astrometric measurement error
  was less than 0.3~mas; the dashed line indicates $\parsym_{\rm
  ast}=\parsym_{\rm phot}$.  In measuring the astrometric parallaxes, we
  held all reference stars but the one labeled fixed at their
  photometric parallaxes, and then solved for the parallax of the
  reference star in question.}
\label{fig:photpar}
\end{figure}

\subsection{Photometry}
\label{sec:photometry}

For the HRC photometry, we measured fluxes in the averaged
flat-fielded images for each epoch (see \S~\ref{sec:hrc}).  Since the
HRC flux for \rxj\ will help constrain the different components of the
optical emission \citep{kvkm+03}, we took care to correct for all
known systematic effects, following the prescription of \citet{sjb+05}
for flat-fielded, non-drizzled images.  Briefly, we first measured
counts in a range of apertures using {\tt DAOPHOT} \citep{stetson87},
and sky in a 50 to 60 pixel annulus (1\farcs33--1\farcs59).  We chose
an aperture of 3 pixels radius (0\farcs08) as a compromise between
adequate signal-to-noise ratio for fainter targets and minimal effects
from PSF variations over the detector.  Next, we corrected for
pixel-area variations and charge transfer inefficiencies following
\citet{adh5.0}.  We used brighter, isolated stars to determine the
aperture correction to an 18-pixel (0\farcs48) aperture, and added a
fixed $-0.103~$mag to correct to ``nominal infinity'' \citep{sjb+05};
the latter includes the small, $-0.003~$mag effect from contamination
of the sky by starlight.  Finally, we added the zero point of 25.623
to obtain magnitudes in the Vega system.

We checked for, but found no significant variations for any star (for
star 113, we rejected epochs 1 and 3, since a trail from star A passed
over the image).  For \rxj, where we list the individual measurements
in Table~\ref{tab:obs}, the standard deviation is 0.10~mag, consistent
with the error of 0.09~mag and with numbers for other faint sources
(e.g., 0.09 and 0.10~mag for stars 112 and 108, respectively).  We
list the average magnitudes in Table~\ref{tab:photpar}.

For STIS, we again used {\tt DAOPHOT} to measure fluxes, choosing a
2-pixel (0\farcs1) aperture.  We corrected for charge transfer
inefficiencies following \citet[][ see also \citealt{gbmak06}]{gk02b},
and used brighter stars to correct to a 10 pixel (0\farcs5) radius
aperture.  To place our magnitudes roughly on the STScI system, we
added an additional $-0.100~$mag to ``nominal infinity,'' and used the
zero point flux of $0.9987\times10^{-19}\mbox{ erg cm}^{-2}\mbox{
s}^{-1}\mbox{ \AA}^{-1}$ for a count rate of $1{\rm~s^{-1}}$, and the
magnitude offset of 21.1, as described in the \hst\ data handbook.  We
stress that while the results are fine for our purpose of inferring
stellar colors (since calibration errors will be taken out), care has
to be taken in interpreting them in terms of absolute fluxes.  In
particular, for the flux of \rxj\ itself, see the detailed discussion
in \citet{kvkm+03}.

For the LRIS photometry, \citet{kvk98} had found that scattered light
from much brighter stars made photometry difficult, and they adopted a
simplified PSF fitting technique.  In order to measure not just
isolated stars, but also ones near others, we modified their procedure
slightly, using the positional information from the HRC and picking a
moffat function instead of a Gaussian as a PSF model.  The results,
listed in Table~\ref{tab:photpar}, are consistent with those of
\citet{kvk98}.  They are also roughly consistent (brighter by
$\sim\!0.03~$mag) with those of \citet{mh98} and \citet{mzh03}.

\section{Photometric parallaxes of background sources}
\label{sec:photpar}
In measuring the parallax of \rxj, we need to worry about bias by
parallaxes of our background sources.  Typical distances may range
from 1 to 10\,kpc, inducing an effect at the 1 to 0.1\,mas level.
Generally, one might expect more nearby sources to bias the fit most,
since these would be brighter and hence have heavier weight.  Given
this, it is good to have distance estimates, which, even if wrong for
individual sources, reduce any systematic bias.  For this purpose, we
use our photometry: we estimate temperature from \bmr\ and flux from
\mr, and then infer a distance assuming the stars are on the main
sequence.

For many stars, we have \mb\ and \mr\ magnitudes from our Keck images
(\S\ref{sec:photometry}).  Some stars, however, were too faint (in
particular in \mb) or too close to brighter stars for reliable
photometry.  For those, we infer \bmr\ and \mr\ from the HRC and STIS
photometry, using empirical color-color relations determined from
stars with uncertainties in \bmr\ smaller than 0.1~mag; see
Fig.~\ref{fig:colors} and Table~\ref{tab:colors}.  We find that all
relations are tight, with root-mean-square residuals of less than
0.05~mag.  Between \hrcbmr\ and \bmr, the relation is almost linear,
while the relations between \hrcbmstis\ and \bmr, and \hrcbmstis\ and
\stismr\ are well-described by quadratic functions.  By way of
verification, we compared our result for \hrcbmr\ and \bmr\ with the
relation expected from synthetic photometry on model atmospheres
\citep{sjb+05}.  As can be seen in Fig.~\ref{fig:colors}, the
agreement is good.

We proceeded by estimating \bmr\ values from \hrcbmr\ and \hrcbmstis\
for all stars; for a given star, we adopt the \bmr\ estimate with the
smallest uncertainty (where the uncertainty includes in quadrature the
measurement uncertainty and the scatter around the required
transformation).  For objects for which \bmr\ was inferred from
\hrcbmstis, we estimated \mr\ from \mstis\ using the relation between
\hrcbmstis\ and \stismr.

In Fig.~\ref{fig:hrd}, we show our adopted \bmr\ and \mr\ estimates in
a color-magnitude diagram.  Also shown are expected colors and
magnitudes for main-sequence stars at various distances, for a
reddening of $A_V=0.25$ (see below).  For these, we used the absolute
magnitudes $M_V$ and colors from \citeauthor{allen} (\citeyear{allen};
specifically, we obtained $M_V$ and $B-V$ from Table 15.7, and
converted to $M_R$ and \bmr\ using Table~15.11; we cannot use Table
15.7 directly since our R-band magnitudes are in the Kron-Cousins
system).  One sees that if our objects are main-sequence stars, they
lie between 2 and 20~kpc; the estimated photometric parallaxes are
listed in Table~\ref{tab:photpar}.  We note that it is very unlikely
that they are not main-sequence stars: if they were giants, they would
be well outside the Galaxy, while if they were white dwarfs, we would
find a significant parallax.

The photometric parallaxes for the brighter stars, which carry most
weight in our astrometry, are all around 0.1~mas, and thus even large
fractional errors in these will not influence our conclusion much.
Nevertheless, it is worth briefly considering the main sources of
error.  Likely, these are the assumed reddening and metallicity.  For
the reddening, we used $E_{B-V}=0.08$, derived by \citet{mh98} from
UBV photometry of stars near \rxj.  This is consistent with the
$E_{B-V}\simeq0.07$ inferred from the observed \bmr\ colors of stars A
and B and their G2 and G5 spectral types (as inferred from the spectra
of \citealt{hmb+97}; see also \citealt{mh98}).  It is somewhat lower
than the total line-of-sight extinction $E_{B-V}=0.14$ estimated from
background infrared emission \citep*{sfd98}.

The above estimates assume solar metallicity.  A sub-solar metallicity
may be more likely, however, since the background sources are
relatively high above the plane and at large Galactocentric radius:
for $(\ell,b)=(244\fdg16,-8\fdg16)$ and at 10~kpc, one finds $z=1.4$
above the Galactic plane and Galactocentric radius $\varpi=16$~kpc,
(we ignored that the plane is warped down by $\sim\!3^\circ$ at
$\ell=240^\circ$; \citealt{mzg+06}).  This may also influence the
reddening: for ${\rm[Fe/H]}=-1.0$, \citet{mh98} infer a lower
reddening $E_{B-V}=0.04$ from their UBV photometry, while from stars A
and B one infers a higher reddening $E_{B-V}\simeq0.14$ (since they
would be intrinsically bluer).

To estimate the maximum effect of sub-solar metallicity, we combined
the increased intrinsic blueness, decreased luminosity, and increased
reddening for star~A.  We found that a change of 0.6~mag in distance
modulus, or a factor 1.3 in photometric parallax.  For stars of later
spectral type, the effect is less.  Since the brighter background
stars are so distant, the additional uncertainty in the parallax of
\rxj\ should be less than 0.1~mas.

\section{Measuring the Parallax}
\label{sec:parallax}

Our goal is to measure the proper motion and parallax of \rxj, but
this has to be done relative to other stars, for which the positions,
proper motions and parallaxes are not known {\em a priori}, and on
images for which our knowledge of the positions, rotations, and scales
is insufficiently accurate.  In our procedure, we fit simultaneously
for our target parameters and all other parameters.  We have to fix
some of those, however, since otherwise the fit is degenerate (e.g., a
net shift between epochs cannot be distinguished from a proper motion
component common to all stars).  We made the following choices.

\begin{figure}
\epsfig{file=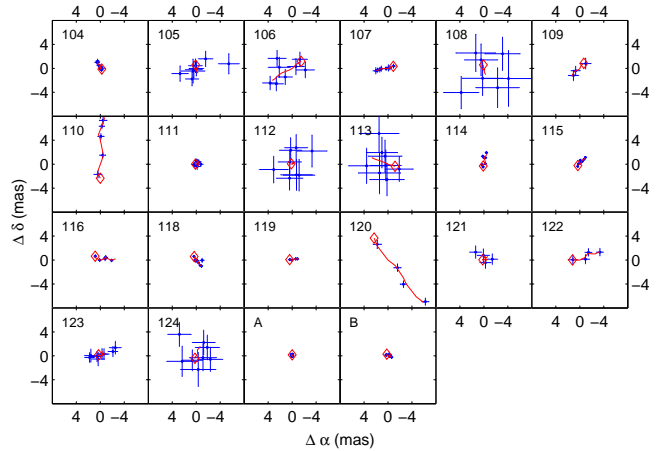,width=0.48\textwidth}
\caption{Positions of the reference stars at every measured epoch,
  with the predicted track based on the best-fit proper motion and
  photometric parallax indicated by the solid line.  The diamond
  indicates the fitted position at the epoch of the first observation.}
\label{fig:pms}
\end{figure}

\begin{deluxetable*}{l c r l l l l l}
\tablecaption{Astrometry of Reference Sources\label{tab:ast}}
\tablewidth{0pt}
\tablehead{&&\colhead{$m_{\rm instr}$}&
  \colhead{$\Delta\alpha$}& \colhead{$\Delta\delta$}&
  \colhead{$\mu_{\alpha}$}& \colhead{$\mu_{\delta}$}&
\\
\colhead{ID}& \colhead{$N_{\rm epoch}$}&\colhead{(mag)}&
\colhead{(arcsec)}&\colhead{(arcsec)}&
\colhead{(mas yr$^{-1}$)}&\colhead{(mas yr$^{-1}$)}&
  \colhead{$\chi^{2}_{\rm red}$}
}
\startdata
X  & 8 &$ -5.7$&\phn\phs$6.1219(8)  $&\phn\phs$1.7026(9)  $&$      -93.9(12) $&\phs$   52.8(13) $&1.13\\
A  & 8 &$-12.0$&\phn\phs$0          $&\phn\phs$0          $&\phn\phs$0       $&\phn\phs$0       $&0.67\\
B  & 8 &$-11.8$&\phn\phs$9.46094(12)$&\phn$   -8.01896(11)$&\phn$   -0.45(17)$&\phn$   -0.18(17)$&0.50\\
104& 7 &$ -9.2$&\phs$   13.92144(16)$&\phn$   -2.65145(15)$&\phn\phs$0.4(2)  $&\phn\phs$0.8(2)  $&0.44\\
105& 8 &$ -6.4$&\phn\phs$4.2079(5)  $&\phn\phs$4.0488(5)  $&\phn$   -0.2(7)  $&\phn$   -0.7(7)  $&1.83\\
106& 8 &$ -6.5$&\phn\phs$3.4296(5)  $&\phn\phs$5.5332(5)  $&\phn\phs$2.9(7)  $&\phn$   -1.8(7)  $&0.93\\
107& 8 &$ -9.3$&\phn\phs$2.03280(16)$&\phs$   10.34588(16)$&\phn\phs$1.7(3)  $&\phn$   -0.3(3)  $&0.51\\
108& 7 &$ -5.4$&\phn\phs$3.8609(14) $&\phs$   11.8761(13) $&\phn$   -0.2(19) $&\phn$   -0.9(18) $&1.19\\
109& 3 &$ -7.8$&\phn\phs$6.1437(5)  $&\phs$   15.4211(5)  $&\phn\phs$1.3(7)  $&\phn$   -1.2(7)  $&1.39\\
110& 5 &$-10.2$&\phn$   -2.9762(3)  $&\phs$   13.0787(3)  $&\phn$   -0.2(4)  $&\phn\phs$6.3(4)  $&0.53\\
111& 7 &$ -9.0$&$      -11.21837(18)$&\phn\phs$7.37302(18)$&\phn$   -0.3(3)  $&\phn\phs$0.3(3)  $&0.98\\
112& 8 &$ -5.8$&\phn$   -1.6390(8)  $&\phn\phs$1.4360(8)  $&\phn$   -0.3(12) $&\phn\phs$0.2(13) $&0.96\\
113& 8 &$ -5.5$&\phn\phs$0.1059(10) $&\phn$   -1.3660(9)  $&\phn\phs$2.3(16) $&\phn\phs$0.9(15) $&0.46\\
114& 7 &$-10.2$&\phn$   -5.84398(9) $&\phn\phs$0.22906(9) $&\phn$   -0.25(14)$&\phn\phs$1.28(14)$&0.99\\
115& 8 &$-10.9$&\phn$   -6.88576(9) $&\phn$   -2.06279(8) $&\phn$   -0.69(13)$&\phn\phs$0.83(13)$&0.62\\
116& 6 &$ -9.8$&$      -10.62105(13)$&\phn$   -1.09834(13)$&\phn$   -1.8(2)  $&\phn$   -0.0(2)  $&0.12\\
118& 8 &$-12.0$&\phn$   -6.52202(9) $&\phn$   -8.86165(9) $&\phn$   -0.68(15)$&\phn$   -0.86(15)$&0.60\\
119& 4 &$-12.8$&$      -10.16052(10)$&$      -10.01889(10)$&\phn$   -0.86(19)$&\phn\phs$0.16(19)$&0.27\\
120& 4 &$ -8.3$&\phn$   -2.8154(4)  $&$      -12.5970(4)  $&\phn$   -4.7(5)  $&\phn$   -6.1(5)  $&0.95\\
121& 4 &$ -7.2$&\phn$   -1.6247(5)  $&$      -13.9458(6)  $&\phn$   -0.4(7)  $&\phn\phs$0.4(7)  $&1.81\\
122& 4 &$ -8.8$&\phn$   -6.5404(4)  $&$      -15.8855(4)  $&\phn$   -2.5(5)  $&\phn\phs$1.0(5)  $&0.19\\
123& 8 &$ -7.0$&\phs$   11.9215(4)  $&\phn$   -7.3533(4)  $&\phn$   -0.6(5)  $&\phn\phs$0.3(5)  $&2.01\\
124& 8 &$ -5.9$&\phs$   12.6335(8)  $&\phn$   -7.6056(8)  $&\phn$   -0.5(11) $&\phn\phs$1.3(11) $&1.00\\

\enddata
\tablecomments{Columns are: ID: star number, with letters following
  \citet{hmb+97}.  Source X is \rxj; for its parallax, see
  Table~\ref{tab:par}. $N_{\rm epoch}$: number of epochs for which a
  source could be measured.  $m_{\rm instr}$: instrumental magnitudes
  defined as $-2.5\log_{10} a$, where $a$ is the average amplitude
  of the ePSF required to fit a given star in a single exposure.
  $\Delta\alpha,\Delta\delta$: position offsets at MJD~52645.1,
  relative to star~A (\S~\ref{sec:absast}), excluding parallactic
  offsets.  $\mu_\alpha,\mu_\delta$: proper motion relative to that of
  star A.  The systematic uncertainty on the proper motion, from
  assuming that star~A has $\mu=0$, is $\sim\!\pmsyst\mbox{ mas
  yr}^{-1}$.  $\chi^2_{\rm red}$: reduced $\chi^2$ values for fitting,
  with degrees of freedom $2N_{\rm epoch}-N_{\rm par}$, where the
  number of parameters $N_{\rm par}=5$ for \rxj, 0 for A, and 4 for
  all other stars (note that these $\chi^2_{\rm red}$ values for
  individual sources are not robust estimates; see
  App.~\ref{app:ast}).  All uncertainties are at the 1$\sigma$ level;
  entries with no uncertainties were fixed at the given values.}
\end{deluxetable*}

First, we chose epoch~3 to have a known plate-scale and position
angle, with the offsets from the header values (Table~\ref{tab:obs})
set by our ground-based absolute astrometry (\S~\ref{sec:absast}).
This sets the absolute scale and orientation of the data.

Second, we also chose epoch~7 to have known plate-scale and position
angle, using the header values with the same offsets as for epoch~3.
This ensures that the scale and position angle do not drift with time
(i.e., it avoids net expansion or net rotation).
We chose this pair of epochs since they are at almost the same
parallactic angle, so their tie cannot influence the parallax.  The
tie will influence the orientation and scale of the proper motions,
leading to an uncertainty of 0\fdg05 and 0.01\% (as estimated from the
differences in orientation and scale found for other epochs below).
This effect, however, is smaller than that of our next step.

\begin{deluxetable*}{c c c c l l l l c}[p]
\tablecaption{Parameters of the Epoch Transformations\label{tab:epoch}}
\tabletypesize{\footnotesize}
\tablewidth{0pt}
\tablehead{
&&\colhead{$x_0$} & \colhead{$y_0$} &
\colhead{$\phi$} & \colhead{$\lambda$} &
\colhead{$\phi_y$}& \colhead{$\lambda_y$}&\\ 
\colhead{Epoch}& \colhead{$N_{\rm star}$}& 
\colhead{(pix)}&\colhead{(pix)}&
\colhead{($10^{-3}\,$deg)}&\colhead{($10^{-5}$)}&
\colhead{($10^{-3}\,$deg)}&\colhead{($10^{-5}$)}&
\colhead{$\chi^2_{\rm red}$}
}
\startdata
 1 &17 &$556.164(3)$&$496.457(3)$&\phs$   99.3(9) $&$      -30.9(12)$&\phn$   -2.9(11)$&\phn$   -1(2)   $&0.30\\
 2 &21 &$521.164(3)$&$444.095(3)$&\phs$   90.5(10)$&$      -33.6(19)$&\phn\phs$8.0(15)$&\phn$   -9(3)   $&0.35\\
 3 &18 &$461.598(3)$&$517.035(2)$&\phs$   17.71   $&$      -34      $&\phn\phs$0      $&\phn\phs$0      $&0.95\\
 4 &21 &$491.148(3)$&$573.595(3)$&\phs$   12.7(9) $&$      -21.0(16)$&\phs$   11.7(13)$&$      -17(2)   $&0.55\\
 5 &17 &$581.178(3)$&$492.582(3)$&\phn\phs$9.1(10)$&$      -29.4(11)$&\phn\phs$1.0(12)$&\phn$   -6(2)   $&1.63\\
 6 &20 &$546.786(3)$&$471.803(3)$&\phs$   14.6(6) $&$      -28.1(12)$&\phs$   14.7(9) $&$      -20.1(16)$&0.85\\
 7 &19 &$467.461(3)$&$520.372(3)$&\phs$   17.71   $&$      -34      $&\phn\phs$0      $&\phn\phs$0      $&0.66\\
 8 &21 &$509.103(4)$&$581.869(4)$&\phs$   10.0(8) $&$      -20.1(16)$&\phs$   16.4(12)$&$      -19(2)   $&0.58\\

\enddata
\tablecomments{Columns are: Epoch: epoch number.  $N_{\rm star}$:
  number of stars included.  $x_0,y_0$: reference position, which is
  the model position of star A in dedistorted coordinates.  $\phi$:
  deviation of the best-fit position angle relative to the nominal
  orientation listed in Table~\ref{tab:obs}. $\lambda$: deviation of
  the best-fit scale relative to the nominal plate scale of
  $28.27{\rm~mas\,\,pixel^{-1}}$.  $\phi_y,\lambda_y$: deviations from
  orthogonality and equality of scale; see Eq.~\ref{eq:tform}.
  $\chi^2_{\rm red}$: reduced $\chi^2$ values for fitting, with
  degrees of freedom $2N_{\rm star}-N_{\rm par}$, where the number of
  parameters $N_{\rm par}=2$ for epochs 3 and~7, and 6 for all others
  (note that these $\chi^2_{\rm red}$ values for individual epochs are
  not robust estimates; see App.~\ref{app:ast}).  All uncertainties
  are at the 1$\sigma$ level; entries with no uncertainties were fixed
  at the given values, which were determined from our absolute
  astrometry (\S~\ref{sec:absast}).}
\end{deluxetable*}

Third, we fix the position of star~A to that determined from the
absolute astrometry, and set its proper motion to 0.  This sets the
reference position and fixes the net proper motion.  Given the proper
motions of other distant stars, we estimate that it introduces an
uncertainty in our net proper motion of $\sim\!\pmsyst~{\rm
mas\,yr}^{-1}$ in each coordinate.  This is very small compared to the
proper motion of \rxj, but comparable to the formal uncertainty, and
thus should be taken into account.  We note that the choice of star A
is arbitrary, and one could have picked another star or an ensemble.
An ensemble might have similar proper motion, however, and our choice
has the advantage of being simple, and of star~A being a good
reference object in that it is bright, has a small photometric
parallax, and is located near the center of the images.

Finally, to fix the mean parallax (which would be indistinguishable
from an epoch-dependent shift in the pointing), we fix the parallaxes
of all stars but our target to the photometric parallaxes from
Table~\ref{sec:photpar}.  To verify whether these were reliable, we
computed solutions where we fitted for the parallax not only of our
target but also of one more star.  Cycling this additional star among
all reference stars, we find good agreement between the photometric
and astrometric parallaxes (see Fig.~\ref{fig:photpar}).  Even the
most deviant points is only $2.3\sigma$ away, and equality has
$\chi^2=22.5$ for 22 degrees of freedom (22 measurements, no free
parameters).  Thus, our photometric parallaxes appear reliable, at
least on average.

\begin{deluxetable}{l c c}
\tablecaption{Motion of \rxj\label{tab:par}}
\tablewidth{0pt}
\tablehead{
\colhead{Parameter} & \colhead{Best-fit Values} & \colhead{$\rho_{\parsym}$\tablenotemark{a}}\\
}
\startdata
$\alpha_{\rm J2000}$\tablenotemark{b}\dotfill & \phs$07^{\rm h}20^{\rm
  m}24\fs9620 \pm 0\fs0009$ & $-0.08$\\
$\delta_{\rm J2000}$\tablenotemark{b}\dotfill &
$-31\degr25\arcmin50\farcs083 \pm 0\farcs011$ & $-0.10$\\
$\mu_{\alpha}$ ($\mbox{mas yr}^{-1}$)\dotfill & $-93.9 \pm 1.2\phn$ & \phs$0.20$\\
$\mu_{\delta}$ ($\mbox{mas yr}^{-1}$)\dotfill & \phs$52.8 \pm 1.3\phn$
& \phs$0.25$\\
$\parsym$ (mas)\dotfill & \phs$2.77 \pm 0.89$ & \nodata\\
\tableline
$D$ (pc)\dotfill & $361^{+172}_{-88}$\\
$\mu$ ($\mbox{mas yr}^{-1}$)\dotfill &\phs$107.8\pm1.2$\\
PA (deg)\dotfill & \phn$-60.6\pm 0.8$\\
$v_{\perp}$ ($\mbox{km s}^{-1}$)\dotfill & $185^{+88}_{-45}$\\
\enddata
\tablecomments{Errors for $\alpha$ and $\delta$ are determined by the
absolute astrometry (\S~\ref{sec:absast}).  Errors for $\mu_{\alpha}$,
$\mu_{\delta}$, and $\parsym$ are formal uncertainties at the
1$\sigma$/68\% confidence level.  There may be additional systematic
uncertainties, of at most \syst~mas for $\parsym$ (see
App.~\ref{app:summary}) and about \pmsyst~$\mbox{mas yr}^{-1}$ for
$\mu_\alpha$ and $\mu_\delta$ (see \S~\ref{sec:parallax}).  Best-fit
values and errors for the other parameters are derived from those for
$\mu_{\alpha}$, $\mu_{\delta}$ and $\parsym$, ignoring the possible
systematic uncertainties.}
\tablenotetext{a}{The correlation coefficient between each parameter
  and the parallax $\parsym$.}
\tablenotetext{b}{At epoch MJD~52645.1 and for equinox J2000,
  excluding any parallactic offset.}

\end{deluxetable}

With all parallaxes fixed except for that of \rxj, we derived the fit
given in Tables~\ref{tab:ast}, \ref{tab:epoch}, and~\ref{tab:par}.
The quality of the fit was good, with $\chi^2=192.1$ for 179~degrees
of freedom (306 data points and 129 parameters), and most stars fit
quite well, as one can see in Fig.~\ref{fig:pms}, where we compare the
measured positions for the reference stars with our fit.  Examining
each epoch independently (Table~\ref{tab:epoch}), we find that only
epoch~5 has $\chi^2_{\rm red}$ significantly greater than unity.
Interestingly, this epoch also had the poorest $\chi^2$ in the
registration (\S~\ref{sec:hrc}).  Looking in detail, we found that the
ePSF fits generally had worse $\chi^2$, and that the small-aperture
instrumental magnitudes were fainter by $\sim\!0.2$~mag compared to
those of the other epochs (this did not affect our photometry, since
the aperture corrections compensated for it).  Visually, the PSF in
the images from epoch~5 appears less sharp, and thus we believe that
this epoch suffered from larger than average focus variations.  We
note, though, that while this explains the fainter small-aperture
magnitudes and poor ePSF fits, it is unclear why the registration or
astrometry should be significantly poorer.  Overall, we felt there was
insufficient reason to reject epoch~5 and it is therefore included in
our final analysis (we examine what happens if we exclude epoch~5 ---
and other epochs --- in App.~\ref{app:jackknife}).

\begin{figure}
\plotone{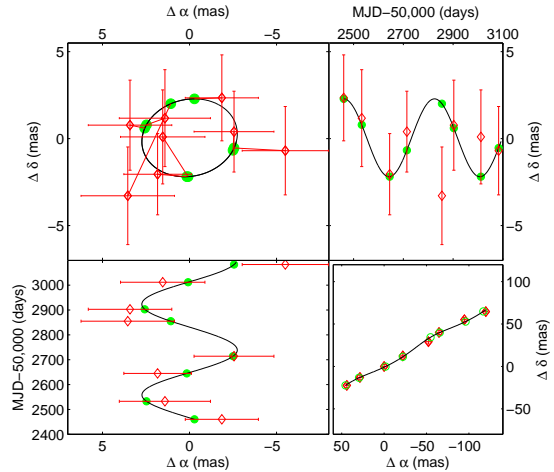}
\caption{Parallax of \rxj.  Upper left: parallactic ellipse showing
  motion in Right Ascension versus that in Declination with the proper
  motion removed.  The filled circles are the predicted positions of
  \rxj, and the open diamonds are the data.  The data are connected to
  the predicted positions at the same epoch.  Upper right: motion in
  Declination versus time with the proper motion removed.  Lower left:
  motion in Right Ascension versus time with the proper motion
  removed.  Lower right: motion in Right Ascension and Declination
  including the proper motion. }
\label{fig:par}
\end{figure}

For \rxj, we find a parallax $\pi=2.77\pm0.89$~mas; the full set of
fitted parameters is listed in Table~\ref{tab:par}, and the fit is
shown in Fig.~\ref{fig:par}.  The fit is good, with $\chi^2_{\rm
red}=1.13$ (for 11 degrees of freedom; 16 measurements and 5
parameters).  From Fig.~\ref{fig:par}, one sees that also for \rxj\
itself epoch~5 fits worse than any of the others, with a net deviation
of $2.4\sigma$, while the next worst is epoch~8 ($1.6\sigma$).  We
return to this in App.~\ref{app:alt}, where we try to estimate
systematic errors in the parallax; we find that these are at most
\syst~mas, i.e., less than half the measurement error.

\ \\
\section{Discussion}
\label{sec:disc}
The distance $d=360^{+170}_{-90}$~pc we measure is roughly consistent
with the estimate of $250\pm25~$pc made by \citet{pph+06} comparing
the hydrogen column density to \rxj\ --- as inferred from fits to its
X-ray spectrum --- with the run of hydrogen column with distance --- as
inferred from sodium columns to stars with known distances.
Conversely, our parallax measurement indicates that the column density
inferred from the spectral fits, which necessarily depends on the form
assumed for the intrinsic spectrum, is reasonable (if perhaps slightly
low; a distance slightly closer than the parallax distance is also
found for \rxjw: \citet{pph+06} find $d=135\pm25$~pc, while our
preliminary parallax yields $167^{+18}_{-15}$~pc).

\begin{figure*}
\plotone{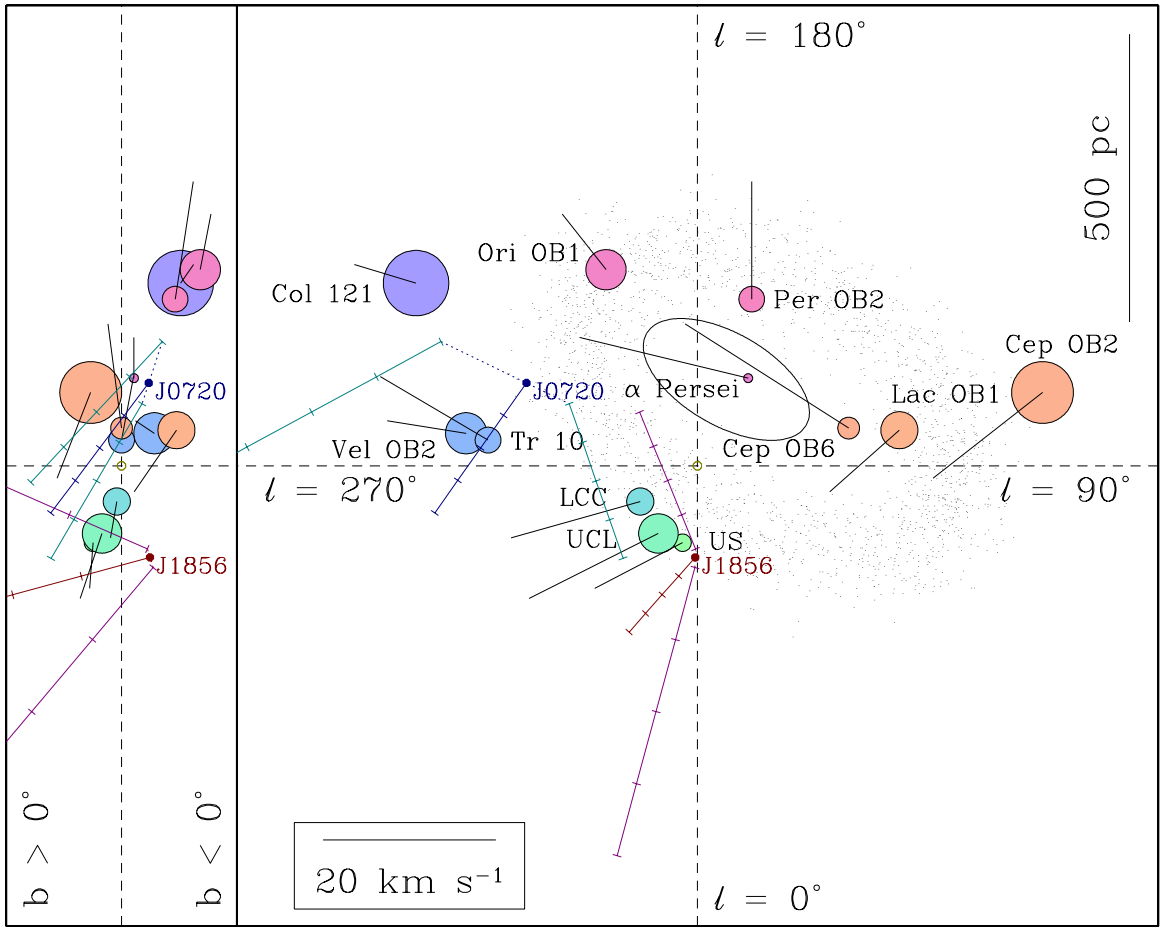}
\caption{Origins of two nearby isolated neutron stars.  This figure,
  adapted from Fig.~29 of \citet{dzhdb+99}, shows a view looking down
  on the Galactic plane (right panel, with Galactic longitude $\ell$
  as indicated) and a view across the plane (left panel, with Galactic
  latitude $b$ as indicated).  The filled circles show kinematically
  detected OB associations (filled circles; \citealt{dzhdb+99}) along
  with their streaming motions, and the scattered dots represent the
  Gould Belt \citep{olano82}.  The present locations of \rxjw\
  (\citealt{kvka02}; van~Kerkwijk et~al.\ 2006, in prep.)  and \rxj\
  (this work) are indicated, as are trajectories back in time for the
  nominal distance and proper motion, assuming radial velocity $v_{\rm
  rad}=0$, as well as using the 1$\sigma$ distance limits and $v_{\rm
  rad}=\pm0.935v_{\perp}$ (see \S~\ref{sec:origin}).  The tick marks
  along the trajectories occur every 0.5~Myr; Galactic acceleration is
  negligible over these timescales.  The scale for velocity of
  $20\mbox{ km s}^{-1}$ for the associations is shown along the
  bottom, while the linear scale of 500~pc is along the right edge.
  For reference, a source with a velocity of $20\mbox{ km s}^{-1}$
  will traverse 500~pc in 25~Myr.}
\label{fig:origin}
\end{figure*}

Our parallax implies that \rxj\ is roughly a factor two more distant
than \rxjw.  This agrees well with the simple estimate of
\citetalias{kvka02}, which was based on the first-order assumptions that
the optical flux for different sources scaled as $F_{\nu} \propto T
(R/d)^2$, that the radii were similar, and that the temperature $T$ in
the region emitting optical photons scaled with the temperature
determined from fits to the X-ray spectra.  Although this agreement
may be a coincidence, it suggests that the mismatch between black-body
fits to the X-ray and optical emission arises because the surface does
not emit like a black body (i.e., that the temperatures in the layers
of the atmosphere emitting X-ray and optical photons are different but
related; e.g., \citealt{mzh03}; \citealt*{ztd04}), and not because the
optical and X-ray emission originate in separate hot and cool regions
(e.g., \citealt{br02}).

While the X-ray spectrum and flux of \rxj\ had been observed to be
relatively constant for years after its discovery (like for the other
INS), observations in 2002 and 2003 showed a surprising spectral
change \citep{dvvmv04,vdvmv04}.  The spectrum hardened significantly,
although the flux stayed relatively constant.  Our ACS observations
occurred during the same time span.  In \S~\ref{sec:photometry}, we
looked for variability in the optical flux of \rxj, but did not find
any.  This may not be entirely unexpected, however, if the optical
flux indeed results from the Rayleigh-Jeans tail of some hot region
(and scales as $R^2 T $, as above).  Over the course of our
observations, the temperature of the X-ray blackbody changed by 6\%,
from $kT=88.3\pm0.3$~keV (2002~November) to $93.8\pm0.4$~keV
(2004~May), while the blackbody angular diameter dropped $\sim\!6$\%
over the same period \citep{htdv+06}.  This implies a decrease in
optical flux of $\sim\!6\%=0.07$~mag, which is smaller than our
photometric uncertainty of 0.09~mag (Table~\ref{tab:obs}).  Even
binning the observations we do not see significant variability:
comparing the mean of epochs 1--4 with 5--8 there is a decrease of
0.08~mag, but the uncertainties on each mean are 0.05~mag.

\subsection{Origin and Age}
\label{sec:origin}
With the distance and proper motion of \rxj, we can estimate the space
motion and try to determine its origin and age.  As the simplest
estimate, we determine the time required for \rxj\ to reach its
current location out of the Galactic plane.  The current Galactic
latitude is $b=-8.16\degr$, and the proper motion in that coordinate
is $\mu_b=-59.3\mbox{ mas yr}^{-1}$.  So, assuming it started at
$b=0$, it has taken $\sim\!b/\mu_b=\expnt{5}{5}$~yr to reach that
height.  However, this value is rather inaccurate.  At a distance of
$360$~pc, \rxj\ is at $z-z_{\odot}=-50.7$~pc, where $z$ is the height
above the Galactic plane, and the Sun is at $z_{\odot}\approx10$~pc
compared to local OB stars \citep{reed97,ecca06}.  So \rxj\ is
actually at $z\approx-40$~pc below the mid-plane.  Given that this is
comparable to the local scale-height of 30--40~pc of OB stars
\citep{reed00,ecca06}, we cannot actually use this method for a useful
age estimate, but can only derive a rough upper limit of
$\sim\!1$~Myr.

We can derive a more meaningful estimate from considering not simply
the scale height of the OB population, but instead the actual
locations of individual associations.  In Fig.~\ref{fig:origin}, we
show the nearby Galactic OB associations (taken from
\citealt{dzhdb+99}), and overdraw the current location and the
previous motion of \rxj\ for the distance corresponding to the
best-fit parallax and zero radial velocity $v_{\rm rad}$.  We also
show the result for distances corresponding to
$\parsym\pm\sigma_\parsym$, and for $v_{\rm rad}=\pm0.935 v_{\perp}$,
where $v_{\perp}=185{\rm~km\,s^{-1}}$ is the velocity in the plane of
the sky inferred from the proper motion and best-fit parallax, and the
numerical factor $\cot(\cos^{-1} 0.683)=0.934$ corresponds to the
expected $1\sigma$ range in $v_{\rm rad}$ for random orientations.  In
our estimates of the space velocity, we corrected for the Sun's motion
relative to the Local Standard of Rest, of
$(U,V,W)_\odot=(10.00,5.25,7.17){\rm~km\,s^{-1}}$ \citep{db98}, as
well as for differential galactic rotation, using Oort constants
$(A,B)=(14.82,-12.37){\rm~km\,s^{-1}\,kpc^{-1}}$ \citep{fw97}.  For
comparison, we also show the location and motion of \rxjw\ (using
$[\mu_\alpha,\mu_\delta]=[0.327,-0.059]{\rm~mas\,yr^{-1}}$ and
$\parsym=6.0\pm0.6{\rm~mas}$ [van Kerkwijk, Kaplan, \& Anderson 2007,
in preparation]).

From Fig.~\ref{fig:origin}, one sees that, as discussed in
\citet{wal01}, \rxjw\ plausibly came from the Upper~Scorpius or Upper
Scorpius Lupus OB association $\sim\!0.4$~Myr ago.  For \rxj, as
suggested by \citet{mzh03} and \citet{kaplan04}, an origin in the
Trumpler~10 (Tr~10) OB association $\sim\!0.7$~Myr ago seems likely.
This is plausible in $\ell$, $b$, and distance, with little radial
velocity required (values between $-20$ and $+50\mbox{ km s}^{-1}$ are
compatible with the nominal distance).  An origin in the Vela~OB2
association is less likely but not impossible; it would imply a
slightly smaller age ($\sim\!0.6$~Myr).  For a substantially larger
radial velocity ($\gsim\!500\mbox{ km s}^{-1}$), origins in the Lower
Centaurus Crux, Upper Centaurus Lupus, or Upper Scorpius OB
association are possible.  Even for those, however, the age would
still be below 1~Myr (as it has to be from the the proper motion in
galactic latitude, see above; Fig.~\ref{fig:origin} shows that an age
in excess of 1~Myr is excluded also for \rxjw).

Overall, among all of the OB associations, an origin in Tr~10 seems
most likely, with \rxj\ approaching within $\sim\!17$~pc of the core
(the radius of the association is $\sim\!30$~pc).  This should be
compared to 70~pc for Vela~OB2, although this is a larger association
with a radius of $\sim\!55$~pc.  Of course, the fact that the
trajectory points back to Tr~10 could be a coincidence.  To estimate
the probability of such a coincidence, we did a simulation where
we assumed proper motions with the observed magnitude but
with random direction on the sky.  For each direction, we computed the
minimum possible separation achieved among all OB associations from
\citet{dzhdb+99}, as well as the required radial velocity.  We find
that we can get \rxj\ to within 25~pc (comparable to the closest
approach to Tr~10) of any OB association in only 5\% of our trials,
and in only 2\% of our trials if we restrict the radial velocity to
$|v_{\rm rad}| \leq 350\mbox{ km s}^{-1}$ (which is almost $2
v_{\perp}$).  If we allow for the fact that some OB associations are
larger than others, we find that in only 10\% of our trials does \rxj\
approach within the nominal radius, and in only 5\% of our trials for
the restricted radial-velocity range.  If we vary also the magnitude
of the proper motion, we obtain similar results.  Therefore, given the
small size of Tr~10 and the closeness of the approach, we consider it
likely that \rxj\ came from Tr~10.

If we accept an origin in Tr~10, the main uncertainty in the age comes
from our distance uncertainty.  To assess this, we drew parallaxes for
\rxj\ from a normal distribution corresponding to our measurement
uncertainty, and determined what radial velocity and age gave the
closest approach to Tr~10.  We found, as expected, a most probable age
of 0.7--0.8~Myr, and a linear relation between parallax and
age\footnote{A linear relation is expected since the pulsar needs to
traverse a certain physical distance between the line of sight and the
cluster, and, for given proper motion, its space velocity
perpendicular to the line of sight depends linearly on the distance.}
such that ${\rm Age} = 0.7\left(\parsym/2.8\mbox{ mas}\right)\mbox{
Myr}$.  We thus infer an $1\sigma$ range in age of 0.5--1.0~Myr.
Larger parallaxes imply more negative radial velocities (approaching
$-100\mbox{ km s}^{-1}$), while small parallaxes require high positive
radial velocities (up to $400\mbox{ km s}^{-1}$ for $\parsym<2$~mas)
to get \rxj\ to the distance of Tr~10 in the short time allowed.

The age we infer for \rxj\ is comparable to simple estimates based on
models of neutron star cooling \citep[e.g.,][]{kkvkm02}, but it
poses a pair of  puzzles.  The first is that the spin-down age for
\rxj\ of 1.9~Myr is a factor of 2--3 larger than either the cooling
age or our kinematic age (see \citealt{kvk05} and \citealt{vkk06}, who also discuss
possible explanations).  In this respect, \rxj\ may not be unique: the
hotter neutron star \object[RX J1308.6+2127]{RX~J1308.6+2127} has a
similarly long spin-down age of 1.5~Myr \citep{kvk05b}.  The second
possible puzzle is that, kinematically, \rxj\ appears older than
\rxjw, while its temperature is higher.  Thus, the sources likely
differ in some other aspect.  The inferred magnetic field strengths
are indeed different ($\sim\!2$ vs. $\lesssim\!1\times10^{13}$~G;
\citetalias{kvka02}; \citealt{kvk05}), but this may not suffice.  Instead, it may
reflect a small difference in mass, to which cooling appears to be
especially sensitive (for a review, see \citealt{yp04}).

\section{Conclusions}
\label{sec:conc}
We have measured the geometric parallax to the neutron star \rxj\ with
\hst\ to a precision of 30\%, with an additional systematic
uncertainty of $\sim\!15$\%.  To our knowledge, at $B=26.6~$mag, \rxj\
is the faintest optical object for which a parallax has been measured.
While the measurement is too uncertain to lead to useful constraints
on the radius (which would also require better model atmospheres), it
helps greatly to set the demographic scale of the INS as a population
relative to that of the radio pulsars, for which parallaxes of this
magnitude have been measured routinely\footnote{See
\url[http://www.astro.cornell.edu/~shami/psrvlb/parallax.html]{http://www.astro.cornell.edu/$\sim$shami/psrvlb/parallax.html}.}
(using Very Long Baseline Interferometry [e.g., \citealt{ccv+04}] for
regular pulsars and ``timing'' parallaxes [e.g., \citealt{vsbb+01}]
for millisecond pulsars).  Furthermore, the astrometry shows that the
space velocities of \rxj\ and \rxjw\ are typical for radio pulsars
(\citealt*{acc02}; \citealt{fgk06}).

The prospect for small improvements of the parallax are good, as a few
measurements with a large time baseline will significantly constrain
the proper motions of \rxj\ and the reference stars, which will
improve the epoch registration, reduce the systematic uncertainties
(App.~\ref{app:summary}), and reduce the parallax uncertainty through
its covariance with the proper motion and reference position
(Table~\ref{tab:par}).  With such measurement, total (statistical plus
systematic) uncertainties around 20\% may be possible.  To do much
better, however, would require further intense observations, and more
than 100 orbits of {\em HST}.

Until a more accurate distance is available, the best constraints will
probably arise if the data are combined with our improved distance for
\rxjw\ (van~Kerkwijk et al.\ 2006, in prep.).  While \rxj\ has the
poorer distance, in other ways the modeling is more constrained, since
we know the spin period (but see \citealt{tm07}), have an estimate of
its dipole magnetic field strength, and can use the variation with
viewing geometry and the broad absorption feature in the X-ray
spectrum \citep{hztb04}.

At present, the most useful aspect of our measurement may be that they
provide the necessary input for placing \rxj\ on a cooling diagram;
previously, this was done with an erroneous spin-down age, which
disagrees with our kinematic age and leads to the object appearing
hotter than predicted \citep[e.g.,][]{plps04}.  Of course, for proper
comparison, we must be careful to construct cooling curves that
reflect what we know about the surface composition (partially-ionized
hydrogen? see \citealt{vkk06}) and magnetic field
($\sim\!\expnt{2}{13}$~G; \citealt{kvk05}) of \rxj.

\acknowledgements We thank an anonymous referee for helpful comments
in shaping this paper.  D.~L.~K.\ was partially supported by a fellowship
from the Fannie and John Hertz Foundation, and M.~H.~v.~K by a
discovery grant from the National Science and Engineering Research
Council.  Further support for this work was provided by the National
Aeronautics and Space Administration through Hubble award GO-09364.
This research made extensive use of ADS, SIMBAD, and Vizier.
Figure~\ref{fig:origin} was adapted from Fig.~29 of \citet{dzhdb+99}
by permission of the authors and the AAS.

{\it Facilities:} \facility{HST (ACS,STIS)}, \facility{Keck:II
  (LRIS)}, \facility{VLT:Antu (FORS1)}

\appendix

\section{Details of the Parallax Fitting}
\label{app:ast}

Our method of determining the parallax from the position measurements
in individual exposures was described briefly in
\S~\ref{sec:parallax}.  Here, we describe it in more detail.  

We measured positions using the effective-PSF (ePSF) technique, as
discussed in \citetalias{ak04}\footnote{For the ePSF model, distortion
solution, and astrometric software described by \citetalias{ak04}, see
\url[http://spacibm.rice.edu/~jay/HRC]{http://spacibm.rice.edu/$\sim$jay/HRC}.}
and \S~\ref{sec:hrc}, and this gives us 64 measurements of up to 23
stars.  In order to determine the parallax of \rxj, it is necessary to
transform and combine those measurements.  We did this in two stages:
(i) combine measurements from different exposures in each epoch; and
(ii) fit the resulting averages with an astrometric model.

The first stage starts with distortion-corrected positions
$(x,y)_{Ees}$ and the associated uncertainties
$(\sigma_x,\sigma_y)_{Ees}$, where $E=1\ldots N_{\rm epoch}$ is the
epoch, $e=1\ldots N_{\rm exp}$ is the exposure, and $s=1\ldots N_{\rm
star}$ is the star number, and the uncertainties are inferred from the
ePSF fit (\S~\ref{sec:hrc}).  For each epoch $E$, we determine average
transformations to a common frame by solving for six parameters
$\{(\Delta x,\Delta y)_{Ee}, \psi_{Ee}, \eta_{Ee}, \psi_{y,Ee},
\eta_{y,Ee}\}$, which relate averaged positions $(x,y)_{Es}$ in the
common frame to the model positions $(\hat x,\hat y)_{Ees}$ in each
exposure, by
\begin{eqnarray}
\hat x_{Ees} &=& (1+\eta_{Ee})\left[(x_{Es}-\Delta x_{Ee})\cos \psi_{Ee}
  + (y_{Es}-\Delta y_{Ee})\sin\psi_{Ee}\right], \nonumber \\
\hat y_{Ees}^\prime& =& (1+\eta_{Ee})\left[-(x_{Es}-\Delta x_{Ee})\sin \psi_{Ee}
  + (y_{Es}-\Delta y_{Ee})\cos\psi_{Ee}\right], \nonumber \\
\hat y_{Ees} &=& (1+\eta_{y,Ee})\left[ -\hat x_{Ees} \sin\psi_{y,Ee}
  + \hat y_{Ees}^{\prime}\cos \psi_{y,Ee}\right].
\label{eq:tformexp}
\end{eqnarray}
Here, $(\Delta x,\Delta y)_{Ee}$ are the offsets of each exposure
compared to the common frame, $\psi_{Ee}$ is the difference in
rotation, $\eta_{Ee}$ is the difference in plate-scale, and
$\psi_{y,Ee}$ and $\eta_{y,Ee}$ are additional parameters that
represent the non-orthogonality of the $x$ and $y$ axes and the
difference between the scales of those axes.  Note that while we use
all six parameters in general, we also made trials with the scale
difference and non-orthogonality fixed to zero; see
App.~\ref{app:combination}.

\begin{figure*}
\plotone{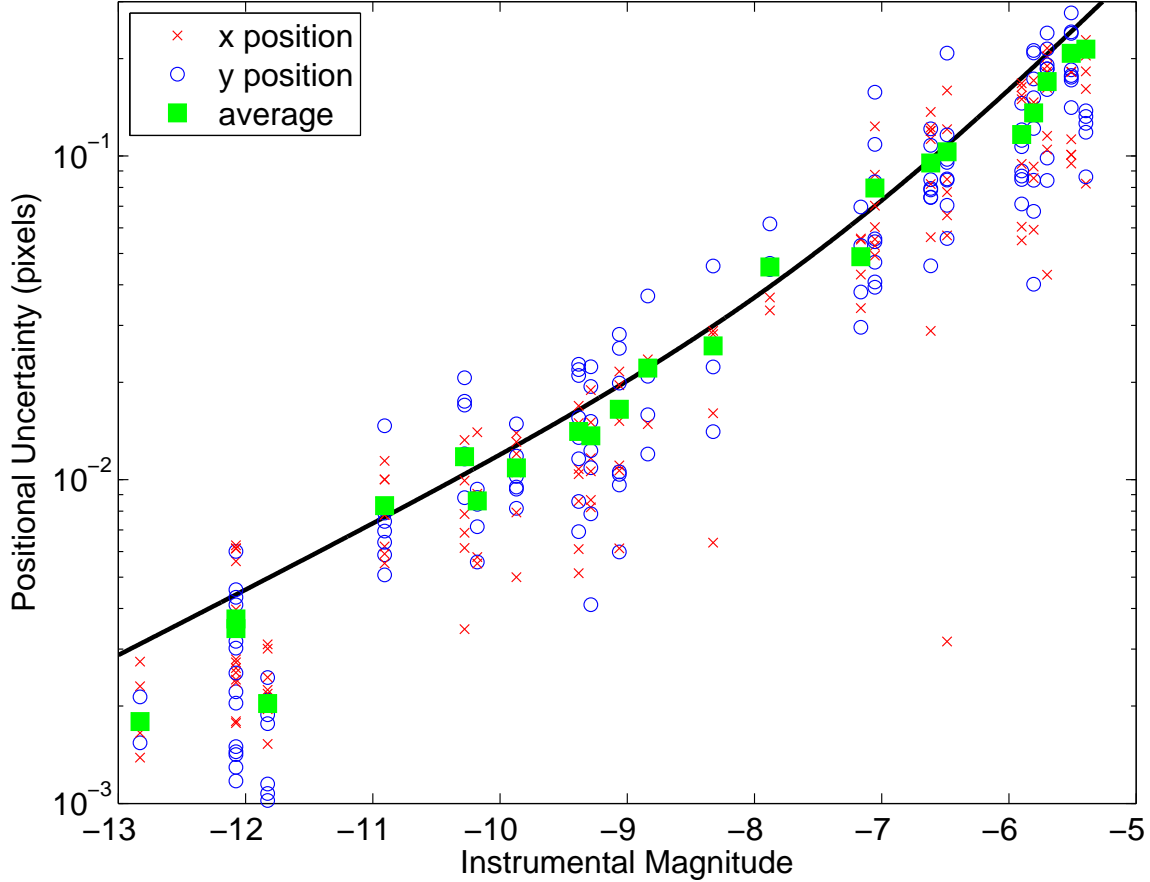}
\caption{Positional uncertainty in pixels versus instrumental
  magnitude for the HRC data---compare to Fig.~1 from
  \citetalias{kvka02} and Fig.~2 from \citetalias{ak04}.  We show the
  standard deviations of the individual positions around the average
  position for each epoch (Eq.~\ref{eq:sd}), in both $x$ (crosses) and
  $y$ (open circles).  We also show the root-mean-square average of
  all the $x$ and $y$ standard deviations (filled squares).  Note that
  for the brightest stars, the standard deviations are expected to
  underestimate the true measurement uncertainties
  (App.~\ref{app:ast}).  The curve gives the relation expected from
  our simulations (Eq.~\ref{eq:sigma}).  The simulations may also
  underestimate the true uncertainty at the brightest magnitudes 
  (App.~\ref{app:uncert}).}
\label{fig:err}
\end{figure*}

In our fits, we take the common frame to be the first exposure, i.e.,
for this exposure all transformation parameters are zero, and one has
$(\hat x,\hat y)_{E1s}=(x,y)_{Es}$.  We solve for the remaining
transformation parameters and the average positions $(x,y)_{Es}$ at
the same time, using a numerical $\chi^2$ minimization routine
(\texttt{mrqmin} from \citealt{numrec}) to minimize,
\begin{equation}
\chi^2 = \sum_{s=1}^{N_{\rm star}} \sum_{e=1}^{N_{\rm exp}} \left[
  \left(\frac{x_{Ees}-{\hat x}_{Ees}}{\sigma_{x,Ees}}\right)^2+
  \left(\frac{y_{Ees}-{\hat y}_{Ees}}{\sigma_{y,Ees}}\right)^2\right].
\label{eq:combchi2}
\end{equation}
To verify whether our fits our reasonable, we use the global $\chi^2$
value, and compute $\chi^2$ values for individual exposures and stars.
For a more direct comparison with our input uncertainties
$\sigma_{Ees}$, we also calculate the standard deviation for
individual sources, 
\begin{equation}
s^2_{x,Es}=\frac{1}{N_{\rm exp}-1} 
           \sum_{e=1}^{N_{\rm exp}} \left({\hat x}_{Ees}-x_{Es}\right)^2.
\label{eq:sd}
\end{equation}
and the same for $y$.  Since $N_{\rm exp}$ is sometimes small for a
given star, the standard deviation is not always a good estimator of
the true uncertainty, but on average we have found that it is quite
reasonable (see \S~\ref{app:uncert} and Fig.~\ref{fig:err}).

In considering $\chi^2$ values and standard deviations for individual
exposures and objects, one has to keep in mind that the transformation
parameters and the average positions are not independent.  In our
case, the four brightest stars (A, B, 118, and 119) dominate the fit.
These provide eight measurements per exposure (4 in each coordinate),
but there are six free transformation parameters, and hence the fit
will partly adjust to remove measurement errors.  As a result, the
$\chi^2$ for these sources are often much less than $2N_{\rm exp}-2$,
the value one would expect if the transformation were independent of
the average positions.  For fainter stars, however, which have much
less effect on the transformation, ones does expect (and we find)
$\chi^2\simeq2N_{\rm exp}-2$.  Of course, for the same reason, the
standard deviations calculated using Eq.~\ref{eq:sd} will
underestimate the true measurement uncertainties for the brightest
stars.

Since we fit for all parameters at the same time, the above
covariances are taken into account automatically.  For instance, for
star 119, our fit yields errors $\sigma_{Es}$ on the average position
similar to the input uncertainty $\sigma_{Ees}$ for a single
measurement; this is a consequence of the fact that substantial
variation in the average position can be compensated for by changes in
the transformation parameters.  In contrast, for fainter stars, we
find $\sigma_{Es}\simeq\sigma_{Ees}/\sqrt{N_{\rm exp}}$, as one would
expect for averaging $N_{\rm exp}$ exposures using a fixed
transformation.  Nevertheless, we worried about our transformation
being dominated by just a few stars, especially since their formal
measurement uncertainties are very small, as low as 0.003~pix for
star~119, at which level systematic effects may well dominate.  Thus,
in App.~\ref{app:uncert}, we verify that our results are robust to
variations in the weight assigned to the brightest stars.

After the combination of the exposures, we have a set of $(x,y)_{Es}$
along with their uncertainties.  We also have a set of times $t_E$
(which determine parallactic offsets $[\Delta \alpha_{\parsym},\Delta
\delta_{\parsym}]_E$ computed using either the JPL DE200 ephemeris or
the approximate formulae of \citealt{allen}, p.\ 670; both methods
gave identical final results) and initial guesses for the position
angle ${\rm PA}_E$ (based on the data headers; Table~\ref{tab:obs})
and plate-scale scale$_{E}$ ($28.27\mbox{ mas pixel}^{-1}$;
\citetalias{ak04}).  In our second stage of the parallax
determination, we want to solve for the full set of stellar parameters
$\mathcal{S}_s$ and epoch transformations parameters $\mathcal{E}_E$.

The stellar parameters $\mathcal{S}_s$ are the reference celestial
position $(\Delta \alpha,\Delta \delta)_s$ of star $s$ relative to
star A at time $t_0$ (note that we chose equal scales for $\Delta
\alpha$ and $\Delta \delta$, so no $\cos\delta$ term is needed), the
proper motion $(\mu_\alpha,\mu_\delta)_s$, and the parallax
$\parsym_s$ (for individual stars, the proper motion or parallax may
be fixed or set to zero).  In terms of these parameters, the celestial
position of star $s$ at epoch $E$ is given by,
\begin{eqnarray}
\Delta \hat \alpha_{Es} & = & \Delta \alpha_s
+ \mu_{\alpha,s}(t_E-t_0) + \parsym_s \Delta\alpha_{\parsym,E},
\nonumber \\
\Delta \hat \delta_{Es} & = & \Delta \delta_s
+ \mu_{\delta,s}(t_E-t_0) + \parsym_s \Delta\delta_{\parsym,E},
\label{eq:ad}
\end{eqnarray}

For the exposures, the parameters $\mathcal{E}_E$ are the position
$(x_0,y_0)_E$ on the detector of the reference position (i.e.,
$[\Delta\alpha,\Delta\delta]=[0,0]$, or the [model] position of
star~A), the difference $\phi_E$ between the initial guess ${\rm
PA}_E$ for the position angle and the fitted value, the difference
$\lambda_E$ between the initial guess ${\rm scale}_E$ for the
plate-scale and the fitted value, and additional parameters
$\phi_{y,E}$ and $\lambda_{y,E}$ that represent the non-orthogonality
of the $x$ and $y$ axes and the difference between the scales of those
axes.  With these parameters, the transformation from the celestial position
of each star at each epoch $(\Delta \hat \alpha,\Delta \hat 
\delta)_{Es}$ to the model positions $(\hat x,\hat y)_{Es}$ is given
by, 
\begin{eqnarray}
\hat x^\prime_{Es} & = & {\rm scale}_E^{-1}(1+\lambda_E) \left[ -\Delta \hat \alpha_{Es} \cos \left({\rm
    PA}_E + \phi_E\right) + \Delta \hat \delta_{Es}\sin\left({\rm PA}_{E} +
    \phi_E\right)\right],\nonumber \\
\hat y^{\prime}_{Es} & = & {\rm scale}_E^{-1}(1+\lambda_E) \left[ \Delta \hat \alpha_{Es} \sin \left({\rm
    PA}_E + \phi_E\right) + \Delta \hat \delta_{Es}\cos\left({\rm PA}_{E} +
    \phi_E\right)\right],\nonumber \\
\hat x_{Es} & = & x_{0,E} +\hat x^\prime_{Es},\nonumber\\
\hat y_{Es} & = & y_{0,E} +(1+\lambda_{y,E}) \left[ -\hat x^\prime_{Es} \sin \phi_{y,E}
    + \hat y^{\prime}_{Es}\cos \phi_{y,E}\right].
\label{eq:tform}
\end{eqnarray}
Note that our transformation uses 6 parameters (as does that of
\citetalias{ak04}), not just the normal four parameters of reference
position, position angle, and scale, but we can also set
$\phi_{y,E}=\lambda_{y,E}=0$ to have a four-parameter transformation
(see App.~\ref{app:combination}).

In the fit, we use Eqs~\ref{eq:ad} and~\ref{eq:tform} to calculate
predicted positions, and fit simultaneously for $\mathcal{E}_{E}$ and
$\mathcal{S}_{s}$ by minimizing,
\begin{equation}
\chi^2 = \sum_{s=1}^{N_{\rm star}} \sum_{E=1}^{N_{\rm epoch}} \left[
  \left(\frac{x_{Es}-{\hat x}_{Es}}{\sigma_{x,Es}}\right)^2+
  \left(\frac{y_{Es}-{\hat y}_{Es}}{\sigma_{y,Es}}\right)^2\right].
\label{eq:parchi2}
\end{equation}
This expression is very similar to the one for the combination of the
exposures (Eq.~\ref{eq:combchi2}), the only difference being that for
the combination the positional parameters (the average positions) were
independent of exposure number, while here there is a dependence on
the time of each epoch (Eq.~\ref{eq:ad}).  Because of this similarity,
our implementation uses the same fitting routine for both steps.

Note that the above method departs from that used by
\citetalias{kvka02}, which solved first for $\mathcal{E}_E$, then
$\mathcal{S}_s$, and iterated.  We found that while our iterative
solution was not biased, direct minimization was more precise, could
more correctly disentangle the effects of stellar proper motions and
uncertainties in $\mathcal{E}_E$, and found more easily the global
$\chi^2$ minimum.

Our fit automatically yields a global $\chi^2$, but, like for the
exposure combination, we can also calculate $\chi^2$ values for each
individual star (Table~\ref{tab:ast}) and each individual epoch
(Table~\ref{tab:epoch}), and these are useful in identifying problems.
We stress again, however, that these are not rigorous $\chi^2$ values,
because the exposure parameters $\mathcal{E}_E$ are common to all of
the stars, and the $\chi^2$ for a single star does not take this into
account (similarly, $\mathcal{S}_s$ are common to all of the epochs,
and the $\chi^2$ for a single epoch does not take this into account).
As for the combinations, this affects the brightest stars in
particular; from Table~\ref{tab:ast}, one indeed sees that stars A, B,
118, and 119 have $\chi^2_{\rm red}$ substantially below unity.

\section{Alternate Schemes for the Parallax Measurement}
\label{app:alt}

We made a number of choices in determining the positions of stars in
individual exposures and in combining these positions measurements.
While we had reasons to prefer some schemes over others, we wished to
verify that our final measurements did not depend on the detailed
choices that we made.  Therefore, we present the consequences of
different ways of measuring positions (\S~\ref{app:pos}), estimating
uncertainties (\S~\ref{app:uncert}), identifying ``bad'' measurements
(\S~\ref{app:rejection}), and combining exposures into averages and
fitting these to an astrometric model (\S~\ref{app:combination}).  We
also discuss a few statistical tests of the robustness of our results
(\S~\ref{app:jackknife}) and end with a summary of the possible
sources of systematic error (\S~\ref{app:summary}).

We note that all of the steps in the analysis (\S~\ref{sec:hrc},
\S~\ref{sec:parallax}, and App.~\ref{app:ast}) were performed
independently by two of us (DLK and MHvK), using separate routines.
We cross-checked the results at many points in the analysis, ensuring
results were identical.

\subsection{Alternate Position Measurements}
\label{app:pos}

At the start of the project, we wished to see if the measurement
scheme of \citetalias{ak04} (Appendix~C) was optimal.  We did a number
of experiments, and settled on a slightly different scheme, described
in \S~\ref{sec:hrc}.  Overall, we found that, for our project, the
choice of scheme had little influence: differences in the final
parallax were less than 0.15~mas, and we find the same parallax also
if we simply use the routines provided by \citetalias{ak04} and do
careful rejection of outliers (App.~\ref{app:rejection}).  However,
the differences may be important for brighter sources, and hence we
briefly describe our experiments below.

First, we varied the way pixels are weighted, taking into account only
Poisson noise, as in \citetalias{ak04}, also read-noise, or even a
small, few percent ``flat-field'' error (proportional to the flux).
We found that using just read-out and Poisson noise provided superior
$\chi^2$ for the combination of exposures.  We also tried determining
weights based on the ePSF model instead of the observed counts.  Here,
using observed counts has the disadvantage that statistical
fluctuation towards low (high) counts get too high (low) a weight),
while using a model has the problem that, due to focus changes, etc.,
the model may not be precise.  The results, however, where not
significantly different.

Second, we experimented with the fit region, trying the
$5\times5$~pixel box used by \citetalias{ak04}, smaller and larger
boxes, as well as circular regions with a range of radii.  Overall, we
found that circular regions gave slightly smaller residuals in
combining exposures.  We picked a radius of 2.56~pixels to minimize
the dependence of the number of pixels in the region on centroid
position (the average number included is 20.6, with a standard
deviation of 0.7).

Third, we tried determining the sky from annuli around sources, like
\citetalias{ak04} do, including it in the fitting process, and fixing
it globally to the header value \texttt{mdrizsky}.  We found only
slight differences; our choice of a constant sky level was based on a
slightly smaller~$\chi^2$.

Fourth, we tried fitting for the ePSF amplitude as well as fixing
these amplitudes using the average photometry and an aperture
correction for each exposure.  It had minimal effect: using fixed
amplitudes gave $\chi^2$ worse by a few percent in the combination and
astrometry.

\subsection{Alternate Uncertainty Schemes}
\label{app:uncert}

We determined uncertainties in the positions from the $\chi^2$
fitting.  To verify these, we simulated the detection of 10,000 stars
of a range of brightnesses, and calculated standard deviations.  This
simulation included the effects of photon noise and read noise (which
dominate for the fainter stars), but did not include any mismatches in
the ePSF (which will dominate for brighter stars).  We find that the
1$\sigma$ uncertainty in each coordinate can be described
approximately by,
\begin{equation}
\sigma(m_{\rm i})=\left[
 \left(C_1 10^{m_{\rm i}/5.0}\right)^2
+\left(C_2 10^{m_{\rm i}/2.5}\right)^2
  \right]^{1/2}~{\rm pixel},
\label{eq:sigma}
\end{equation}
where $m_{\rm i}=-2.5\log N$ with $N$ the counts in a $5\times5$~pixel
box (or, equivalently, the ePSF amplitude from the fit), and
$C_1\simeq1.14$ and $C_2\simeq36$ are constants determined from the
simulation.  This relation roughly matches expectations, since the
uncertainty should scale as the full-width at half-maximum (FWHM)
divided by the signal-to-noise ratio (SNR), where the SNR has
contributions from Poisson noise ($\propto \sqrt{N}$; first term,
which dominates for bright stars) and read-noise ($\propto N/{\rm
RN}$; second term, which dominates for faint stars; here, ${\rm RN}$
is the read-noise in the pixels covered by the PSF).  With a gain of
unity for flat-fielded images and a FWHM close to 1~pixel, one expects
$C_1\simeq1$, as we find.  Similarly, from the read-noise of
$\sim\!6{\rm~DN}$ per pixel, and considering that 20 pixels are used,
one roughly reproduces~$C_2$.  In Fig.~\ref{fig:err}, we compare the
predictions with the standard deviations found after combination of
the exposures (Eq.~\ref{eq:sd}), and find good agreement.


For the brightest stars, one might expect that systematic effects to
start to dominate, but from neither simulations nor observations do we
find evidence for this (Fig.~\ref{fig:err}), in contrast to what
\citetalias{kvka02} found for WFPC2 data, and what is found for HRC by
\citetalias{ak04}.  For the observations, however, one has to keep in
mind that, for the brightest stars, the standard deviations
underestimate the measurement uncertainty, since part of the
measurement error has been absorbed in the transformations required to
put the exposures on a single reference frame (App.~\ref{app:ast}).
Hence, any systematic error might be hidden in the transformations.

To test the effect of possible systematic errors, we added, in
quadrature, constant terms of up to 0.01~pixel to our input
uncertainties, thus reducing the weight carried by the brightest
stars.  As expected, the overall $\chi^2$ for the exposure combination
and the astrometry decreases, from $\chi^2_{\rm
red,comb}=\sum_E\chi^2_E/\sum_E N_{{\rm dof},E}=1.21$ and $\chi^2_{\rm
red,par}=1.07$ to 0.89 and 0.81, respectively.  The change in
parallax, however, is only 0.005~mas.  We can use these results to
estimate the size of the systematic errors, by determining for what
additional systematic error one obtains $\chi^2_{\rm red}\simeq1$.  We
find this is at about 0.007~pix, consistent with the results of
\citetalias{ak04}.


\subsection{Alternate Rejection Schemes}
\label{app:rejection}
During our analysis, we found that a very important part in obtaining
a good registration is the rejection of measurements that are biased
by a bad pixel or a cosmic ray.  The scheme we used was meant to be as
objective as possible, being based on identifying bad pixels in an
entirely automatic way.  Yet, we still needed to reject some
additional measurements (\S~\ref{sec:hrc}).  As an alternative, we
identified bad measurements from large deviations of the positions
and/or instrumental magnitudes of single observations from the
average.  We did not employ a formal rejection criterion but examined
all of the data manually, iteratively registering the exposures and
removing outliers.  For the brightest four stars, very few exposures
were rejected, while for the fainter ones we included on average
7~exposures per epoch.  With this set, we derived a parallax very
similar to our best value.

\subsection{Alternate Combination Schemes}
\label{app:combination}

We combine our individual measurements into averages using a
six-parameter transformation between exposures, and again use a
six-parameter transformation in our astrometric solution.  Although
these choices are grounded in the work of \citetalias{ak04}, we tried
some alternatives: (i) combining exposures with two or four-parameter
transformations; (ii) tying the epochs using a four-parameter
transformation; and (iii) instead of merging the exposures, treating
each exposure as a separate epoch, and fit for 64 transformations
$\mathcal{E_E}$ simultaneously.

We found that a different choice of combining the exposures did not
change the final parallax significantly (i.e., by more than 0.1~mas),
although it did affect the quality of the fit: with fewer parameters,
the $\chi^2$ values were significantly higher.  Using only 4
parameters for the astrometric tie has a more drastic effect on the
quality of the fit.  The parallax, however, decreased by only 0.1~mas
compared to our main result.

When we treated all exposures as separate epochs, we found a parallax
identical to within 0.01~mas with that derived using our two-stage
approach.  The fit was formally less good than our two-stage result,
which may reflect an underestimate of the uncertainties of bright
stars and/or remaining position measurements contaminated by cosmic
rays.  We made a number of experiments to address these issues, but,
fortunately, the effects of all of them on the parallax is small,
$<0.1$~mas.

\subsection{Statistical Tests}
\label{app:jackknife}

As a statistical measure of the robustness of our result, we tried to
``jackknife'' our data, removing either one star or one epoch from the
fit.  Removing a single star does not change the fit appreciably,
showing that no single star dominates the fit.  On average, there was
no change in the parallax, and the dispersion was only 0.03~mas.  The
dispersion is dominated by one star, B: with B omitted, the parallax
decreased by 0.1~mas (note that B is also the most discrepant point in
Fig.~\ref{fig:photpar}).  Given how small this change is compared to
the statistical uncertainty, we are satisfied with this test.

Removing an epoch had a more dramatic effect.  As can be seen from
Fig.~\ref{fig:par}, some epochs fit significantly better than others.
Specifically, removing epoch~5 improved the quality of the fit
($\chi^2_{\rm red,par}=0.93$ for the overall transformation), and
increased the parallax to $3.32\pm0.96$~mas.  In contrast, removing
epoch~8 gave a less significant change in the quality of the fit
($\chi^2_{\rm red,par}=1.03$), but a larger change in parallax, to
$1.67\pm1.01$~mas.  These were the most extreme examples: the mean
shift was 0.08~mas, and the standard deviation was 0.5~mas (0.3~mas
excluding the epoch~8 result).  Part of the problem with removing
epoch~8 seems to be a covariance with the proper motion: while
removing the other epochs changed the proper motion by $<\!1\sigma$,
removing epoch~8 changed $\mu_{\alpha}$ to $-91.9\pm1.5\mbox{ mas
yr}^{-1}$ (the correlation coefficients between $\parsym$ and
$\mu_\alpha$ or $\mu_\delta$ are significant; see
Table~\ref{tab:par}). 


Finally, we also removed individual exposures when we fit using all of
the exposures without combination (App.~\ref{app:combination}).
Again, the mean of the parallaxes was identical to that in
Table~\ref{tab:par}, and the standard deviation was 0.1~mas.  This
leads to an uncertainty of 0.9~mas for the parallax (following
\citealt{et93}), which is the same as what we derived from error
propagation.

\subsection{Estimate of the Systematic Uncertainties}
\label{app:summary}
The biggest source of systematic uncertainties that could affect our
results are the epochs that we include.  While the other choices, as
described above, led to parallax differences of $<\!0.1$~mas, the
jackknife tests on epochs gave larger changes
(App.~\ref{app:jackknife}).  Overall, we believe our result from
Table~\ref{tab:par} to be reliable, but we should add a systematic
uncertainty to the statistical uncertainty quoted there.  The
magnitude of this uncertainty should be similar to the variation in
the parallax that we see, or \syst~mas.  This may be an overestimate,
as our exposure jackknife tests indicated that the bias was small, but
we prefer to err on the conservative side.  A small number of
measurements in the future that will have a long enough time baseline
to make the proper motion better determined should reduce this
uncertainty significantly.



\end{document}